\documentclass[11pt]{article}
\usepackage{graphicx}
\usepackage{amsmath}
\usepackage{amssymb}
\usepackage{caption2}
\begin{document}
\bibliographystyle{prsty}
\begin{center}
{\large {\bf \sc{  Analysis of the $Y(4140)$  and related  molecular states with  QCD sum rules  }}} \\[2mm]
Zhi-Gang Wang \footnote{E-mail,wangzgyiti@yahoo.com.cn.  }, Zhi-Cheng Liu, Xiao-Hong Zhang     \\
 Department of Physics, North China Electric Power University,
Baoding 071003, P. R. China
\end{center}

\begin{abstract}
In this article, we assume that there exist  scalar ${D}^\ast {\bar
{D}}^\ast$, ${D}_s^\ast {\bar {D}}_s^\ast$, ${B}^\ast {\bar
{B}}^\ast$ and ${B}_s^\ast {\bar {B}}_s^\ast$ molecular states, and
study their masses using the QCD sum rules.  The numerical results
indicate that the masses are about $(250-500)\,\rm{MeV}$ above the
corresponding ${D}^\ast -{\bar {D}}^\ast$, ${D}_s^\ast -{\bar
{D}}_s^\ast$, ${B}^\ast -{\bar {B}}^\ast$ and ${B}_s^\ast -{\bar
{B}}_s^\ast$ thresholds, the $Y(4140)$ is unlikely  a scalar
${D}_s^\ast {\bar {D}}_s^\ast$ molecular state.  The scalar $D^\ast
{\bar D}^\ast$, $D_s^\ast {\bar D}_s^\ast$, $B^\ast {\bar B}^\ast$
and $B_s^\ast {\bar B}_s^\ast$ molecular states maybe not exist,
while the scalar ${D'}^\ast {\bar {D'}}^\ast$, ${D'}_s^\ast {\bar
{D'}}_s^\ast$, ${B'}^\ast {\bar {B'}}^\ast$ and ${B'}_s^\ast {\bar
{B'}}_s^\ast$ molecular states maybe exist.
\end{abstract}



\section{Introduction}

Recently the CDF Collaboration   observed   a narrow structure
(which is denoted as the $Y(4140)$  now) near the $J/\psi\phi$
threshold with statistical significance in excess of 3.8 standard
deviations in exclusive $B^+\to J/\psi\phi K^+$ decays produced in
$\bar{p} p $ collisions at $\sqrt{s}=1.96 \,\rm{TeV}$
\cite{CDF0903}. The mass and  width of the structure are measured to
be $4143.0\pm2.9\pm1.2\,\rm{ MeV}$ and $11.7^{+8.3}_{-5.0}\pm3.7\,
\rm{MeV}$, respectively. The narrow structure  $Y(4140)$ is very
similar to the charmonium-like state $Y(3930)$ near the $J/\psi
\omega$ threshold \cite{Belle2005,Babar2008}. The mass and width of
the $Y(3930)$ are $3914.6^{+3.8}_{-3.4}\pm 2.0 \,\rm{MeV}$ and
$34^{+12}_{-8}\pm 5\,\rm{MeV}$,  respectively \cite{Babar2008}.

There have been several explanations for the nature of the   narrow
structure $Y(4140)$, such as a  $D_s^\ast {\bar D}_s^\ast$ molecular
state \cite{LiuMolecule,Mahajan0903,BranzMolecule,NielsenMolecule,
LiuMolecule2,DingMolecule,HuangMolecule}, an exotic
($J^{PC}=1^{-+}$) hybrid charmonium \cite{Mahajan0903}, a $c \bar c
s \bar s$ tetraquark state \cite{Stancu4},
 or the effect of the  $J/\psi \phi$ threshold  \cite{NoResonance}.

The mass is a fundamental parameter in describing a hadron, in order
to identify  the $Y(4140)$ as a scalar molecular state, we must
prove that its mass  lies  in the region $(4.1-4.2)\, \rm{GeV}$. In
Ref.\cite{Wang0903}, we assume that there exists  a scalar
$D_s^*\bar{D}_s^*$ molecular state in the $J/\psi\phi$ invariant
mass distribution, and study its mass using  the QCD sum rules. The
numerical result indicates that the mass is about
$M_{Y}=(4.43\pm0.16)\,\rm{GeV}$, which
 is  inconsistent with the experimental data.
    The $D_s^\ast {\bar
D}_s^\ast$  is probably  a virtual state and not related to the
meson $Y(4140)$ \cite{Wang0903}. In this article, we extend our
previous work to study the ${D}^\ast {\bar {D}}^\ast$, ${D}_s^\ast
{\bar {D}}_s^\ast$, ${B}^\ast {\bar {B}}^\ast$ and ${B}_s^\ast {\bar
{B}}_s^\ast$ molecular states in a systematic way considering the
$SU(3)$ symmetry and the heavy quark symmetry.

 In the QCD sum rules, the operator product expansion is used to expand
the time-ordered currents into a series of quark and gluon
condensates which parameterize the long distance properties of  the
QCD vacuum. Based on the quark-hadron duality, we can obtain copious
information about the hadronic parameters at the phenomenological
side \cite{SVZ79,Reinders85}.

The article is arranged as follows:  we derive the QCD sum rules for
the scalar ${D}^\ast {\bar {D}}^\ast$, ${D}_s^\ast {\bar
{D}}_s^\ast$, ${B}^\ast {\bar {B}}^\ast$ and ${B}_s^\ast {\bar
{B}}_s^\ast$ molecular states in section 2; in section 3, numerical
results and discussions; section 4 is reserved for conclusion.

\section{QCD sum rules for  the ${D}^\ast {\bar {D}}^\ast$, ${D}_s^\ast
{\bar {D}}_s^\ast$, ${B}^\ast {\bar {B}}^\ast$ and ${B}_s^\ast {\bar
{B}}_s^\ast$ molecular states }
 In the following, we write down  the
two-point correlation functions $\Pi_{J/\eta}(p)$  in the QCD sum
rules,
\begin{eqnarray}
\Pi_{J/\eta}(p)&=&i\int d^4x e^{ip \cdot x} \langle
0|T\left\{J/\eta(x)J/\eta^{\dagger}(0)\right\}|0\rangle \, , \\
J(x)&=&\bar{Q}(x)\gamma_\mu s(x) \bar{s}(x)\gamma^\mu Q(x) \, , \\
\eta(x)&=&\bar{Q}(x)\gamma_\mu u(x) \bar{d}(x)\gamma^\mu Q(x) \, ,
\end{eqnarray}
where $Q=c,b$. We choose  the scalar currents $J(x)$ and $\eta(x)$
to interpolate the molecular states ${D}^\ast {\bar {D}}^\ast$,
${D}_s^\ast {\bar {D}}_s^\ast$, ${B}^\ast {\bar {B}}^\ast$ and
${B}_s^\ast {\bar {B}}_s^\ast$, respectively.

We can insert  a complete set of intermediate hadronic states with
the same quantum numbers as the current operators $J(x)$ and
$\eta(x)$ into the correlation functions $\Pi_{J/\eta}(p)$  to
obtain the hadronic representation \cite{SVZ79,Reinders85}. After
isolating the ground state contributions from the pole terms of  the
scalar  molecular states $Y$ (we use the $Y$ to denote the scalar
molecular states ${D}^\ast {\bar {D}}^\ast$, ${D}_s^\ast {\bar
{D}}_s^\ast$, ${B}^\ast {\bar {B}}^\ast$ and ${B}_s^\ast {\bar
{B}}_s^\ast$), we get the following result,
\begin{eqnarray}
\Pi_{J/\eta}(p)&=&\frac{\lambda_{Y}^2}{M_{Y}^2-p^2} +\cdots \, \, ,
\end{eqnarray}
where the pole residue (or coupling) $\lambda_Y$ is defined by
\begin{eqnarray}
\lambda_{Y} &=& \langle 0|J/\eta(0)|Y(p)\rangle \, .
\end{eqnarray}

 In the following, we briefly outline  the operator product
expansion for the correlation functions $\Pi_{J/\eta}(p)$  in
perturbative QCD. The calculations are performed at  the large
space-like momentum region $p^2\ll 0$. We write down the "full"
propagators $S_{ij}(x)$ and $C_{ij}(x)$ of a massive quark in the
presence of the vacuum condensates firstly \cite{Reinders85},
\begin{eqnarray}
S_{ij}(x)&=& \frac{i\delta_{ij}\!\not\!{x}}{ 2\pi^2x^4}
-\frac{\delta_{ij}m_s}{4\pi^2x^2}-\frac{\delta_{ij}}{12}\langle
\bar{s}s\rangle +\frac{i\delta_{ij}}{48}m_s
\langle\bar{s}s\rangle\!\not\!{x}-\frac{\delta_{ij}x^2}{192}\langle \bar{s}g_s\sigma Gs\rangle\nonumber\\
 &&+\frac{i\delta_{ij}x^2}{1152}m_s\langle \bar{s}g_s\sigma
 Gs\rangle \!\not\!{x}-\frac{i}{32\pi^2x^2} G^{ij}_{\mu\nu} (\!\not\!{x}
\sigma^{\mu\nu}+\sigma^{\mu\nu} \!\not\!{x})  +\cdots \, ,
\end{eqnarray}
\begin{eqnarray}
C_{ij}(x)&=&\frac{i}{(2\pi)^4}\int d^4k e^{-ik \cdot x} \left\{
\frac{\delta_{ij}}{\!\not\!{k}-m_Q}
-\frac{g_sG^{\alpha\beta}_{ij}}{4}\frac{\sigma_{\alpha\beta}(\!\not\!{k}+m_Q)+
(\!\not\!{k}+m_Q)\sigma_{\alpha\beta}}{(k^2-m_Q^2)^2}\right.\nonumber\\
&&\left.+\frac{\pi^2}{3} \langle \frac{\alpha_sGG}{\pi}\rangle
\delta_{ij}m_Q \frac{k^2+m_Q\!\not\!{k}}{(k^2-m_Q^2)^4}
+\cdots\right\} \, ,
\end{eqnarray}
where $\langle \bar{s}g_s\sigma Gs\rangle=\langle
\bar{s}g_s\sigma_{\alpha\beta} G^{\alpha\beta}s\rangle$  and
$\langle \frac{\alpha_sGG}{\pi}\rangle=\langle
\frac{\alpha_sG_{\alpha\beta}G^{\alpha\beta}}{\pi}\rangle$, then
contract the quark fields in the correlation function $\Pi_J(p)$
with Wick theorem, and obtain the result:
\begin{eqnarray}
\Pi_J(p)&=&i\int d^4x e^{ip \cdot x}   Tr\left[\gamma_\mu
S_{ij}(x)\gamma_\alpha C_{ji}(-x) \right] Tr\left[\gamma^\mu
C_{mn}(x)\gamma^\alpha S_{nm}(-x) \right] \, ,
\end{eqnarray}
where the $i$, $j$, $m$ and $n$ are color indexes. Substitute the
full  $s$ and $Q$ quark propagators into the correlation function
$\Pi_J(p)$ and complete  the integral in the coordinate space, then
integrate over the variables in the momentum space, we can obtain
the correlation function $\Pi_J(p)$ at the level of the quark-gluon
degrees  of freedom. The correlation function $\Pi_\eta(p)$ is
calculated in the same way, we prefer neglect the technical details.

In the QCD sum rules for the tetraquark states (irrespective of the
diquark-antidiquark type and the molecule type) which have one or
two heavy quarks, we always  calculate the light quark parts of the
correlation functions in the coordinate-space where  the masses of
the $u$, $d$, $s$ quarks are taken as  small quantities  and treated
perturbatively, and use the momentum-space expression for the heavy
quark propagators \cite{Reinders85}, then transform the resulting
light-quark parts to the momentum-space with $D$-dimensional Fourier
transform \cite{4quarkCalc1,4quarkCalc2,Wang0807}.  The main
uncertainties in the QCD calculations originate from the high
dimensional vacuum condensates, which are known poorly compared with
the low dimensional  vacuum condensates, for example, the quark
condensate $\langle\bar{q}q\rangle$ and the gluon condensate
$\langle \frac{\alpha_s GG}{\pi}\rangle$.

In this article, we carry out the operator product expansion to the
vacuum condensates adding up to dimension-10 and
 take the assumption of vacuum saturation for the  high
dimensional  vacuum condensates, they  are always
 factorized to lower condensates with vacuum saturation in the QCD sum
 rules, and   factorization works well in  large $N_c$ limit.
In the real world,
  $N_c=3$, there are deviations from the factorable formula, we can introduce a factor
  $\kappa$ to parameterize the deviations, for example,
  \begin{eqnarray}
\langle \bar{s} s \rangle^2 \, , \, \langle \bar{s} s \rangle
\langle \bar{s}g_s\sigma G s \rangle \, , \,  \langle
\bar{s}g_s\sigma G s \rangle^2 &\rightarrow& \kappa\langle \bar{s} s
\rangle^2 \, , \, \kappa\langle \bar{s} s \rangle \langle
\bar{s}g_s\sigma G s \rangle \, , \,  \kappa\langle \bar{s}g_s\sigma
G s \rangle^2 \, .
  \end{eqnarray}
In Ref.\cite{Wang0903}, we study the mass $M_{D_s^\ast {\bar
D}_s^\ast}$ with  variation of the parameter $\kappa$ at the
interval $\kappa=0-2$. At the range   $M^2=(2.6-3.0)\,\rm{GeV}^2$,
the value $\kappa=1\pm1$ leads to an uncertainty about
$50\,\rm{MeV}$, which is too small to smear the discrepancy between
the theoretical  prediction and the experimental data.  If we assume
the  $\kappa$ has the typical uncertainty of the QCD sum rules, say
about
 $30\%$, the correction   is rather mild, we
 can neglect the uncertainty safely and take $\kappa=1$, i.e. the
 factorization works well.
   In the QCD sum rules for the masses of
the $\rho$ meson and the
 nucleon,  the value of the $\kappa$ is always larger than $1$ \cite{Leinweber97}. In calculations, we
 observe that  larger  $\kappa$ means slower convergence in the operator
  product expansion, requires larger
 threshold parameters, and results in larger ground state masses. We
 can draw the conclusion tentatively  that the uncertainties (i.e.  $\kappa>1$) in the
 QCD calculations  enlarge the discrepancy between
the theoretical  prediction and the experimental data, our
predictions based on the value $\kappa=1$ are reasonable.

The contributions  from the gluon condensates  are suppressed by
large denominators and would not play any significant roles for the
light tetraquark states \cite{Wang1,Wang2}, the heavy tetraquark
state \cite{Wang0807} and the  heavy molecular state
\cite{Wang0903}. In this article, we take into account them
  for completeness  although their contributions are rather
small.

Once analytical results are obtained,   then we can take the
quark-hadron duality and perform Borel transform  with respect to
the variable $P^2=-p^2$, finally we obtain  the following two sum
rules for the interpolating current $J(x)$:
\begin{eqnarray}
\lambda_{Y}^2 e^{-\frac{M_Y^2}{M^2}}= \int_{4(m_Q+m_s)^2}^{s_0} ds
\rho(s) e^{-\frac{s}{M^2}} \, ,
\end{eqnarray}
where
\begin{eqnarray}
\rho(s)&=&\rho_{0}(s)+\rho_{\langle
\bar{s}s\rangle}(s)+\left[\rho^A_{\langle
GG\rangle}(s)+\rho^B_{\langle GG\rangle}(s)\right]\langle
\frac{\alpha_s GG}{\pi}\rangle+\rho_{\langle \bar{s}s\rangle^2}(s)
\, ,
\end{eqnarray}
the lengthy  expressions of the spectral densities $\rho_0(s)$,
$\rho_{\langle \bar{s}s\rangle}(s)$, $\rho^A_{\langle
GG\rangle}(s)$, $\rho^B_{\langle GG\rangle}(s)$ and $\rho_{\langle
\bar{s}s\rangle^2}(s)$ are presented in the appendix. With a simple
replacement,
\begin{eqnarray}
m_s\, , \,\langle \bar{s} s \rangle \, , \,  \langle
\bar{s}g_s\sigma G s \rangle  &\rightarrow& m_q\, , \,\langle
\bar{q} q \rangle \, , \,  \langle \bar{q}g_s\sigma G q \rangle \, ,
  \end{eqnarray}
we can obtain the corresponding two sum rules for the scalar current
$\eta(x)$.

 Differentiating  Eq.(10) with respect to  $\frac{1}{M^2}$, then eliminate the
 pole residues $\lambda_{Y}$, we can obtain two sum rules for
 the masses of the molecular states $Y$,
 \begin{eqnarray}
 M_Y^2= \frac{\int_{4(m_Q+m_s)^2}^{s_0} ds
\frac{d}{d \left(-1/M^2\right)}\rho(s)e^{-\frac{s}{M^2}}
}{\int_{4(m_Q+m_s)^2}^{s_0} ds \rho(s)e^{-\frac{s}{M^2}}}\, ,
\end{eqnarray}
the corresponding two sum rules for  the scalar current $\eta(x)$
can be obtained analogously.

\section{Numerical results and discussions}
The input parameters are taken to be the standard values $\langle
\bar{q}q \rangle=-(0.24\pm 0.01\, \rm{GeV})^3$, $\langle \bar{s}s
\rangle=(0.8\pm 0.2)\langle \bar{q}q \rangle$, $\langle
\bar{s}g_s\sigma G s \rangle=m_0^2\langle \bar{s}s \rangle$,
$m_0^2=(0.8 \pm 0.2)\,\rm{GeV}^2$, $\langle \frac{\alpha_s
GG}{\pi}\rangle=(0.33\,\rm{GeV})^4 $, $m_q\approx0$,
$m_s=(0.14\pm0.01)\,\rm{GeV}$, $m_c=(1.35\pm0.10)\,\rm{GeV}$ and
$m_b=(4.7\pm0.1)\,\rm{GeV}$ at the energy scale  $\mu=1\, \rm{GeV}$
\cite{SVZ79,Reinders85,Ioffe2005}.

 In the conventional QCD sum
rules \cite{SVZ79,Reinders85}, there are two criteria (pole
dominance and convergence of the operator product expansion) for
choosing  the Borel parameter $M^2$ and threshold parameter $s_0$.
We impose the two criteria on the molecular states to choose the
Borel parameter $M^2$ and threshold parameter $s_0$.

\begin{figure}
 \centering
 \includegraphics[totalheight=5cm,width=6cm]{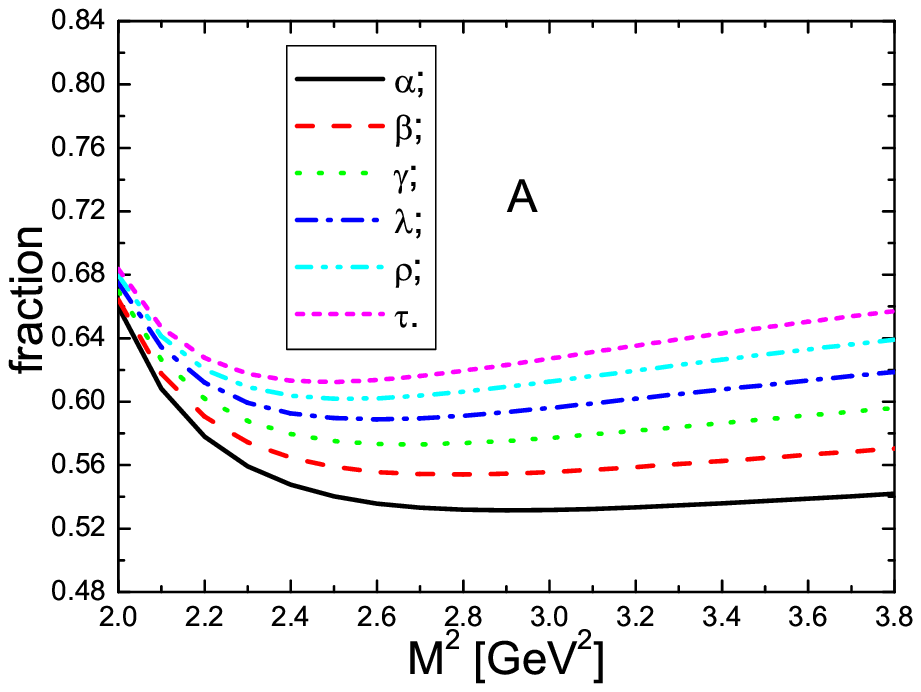}
 \includegraphics[totalheight=5cm,width=6cm]{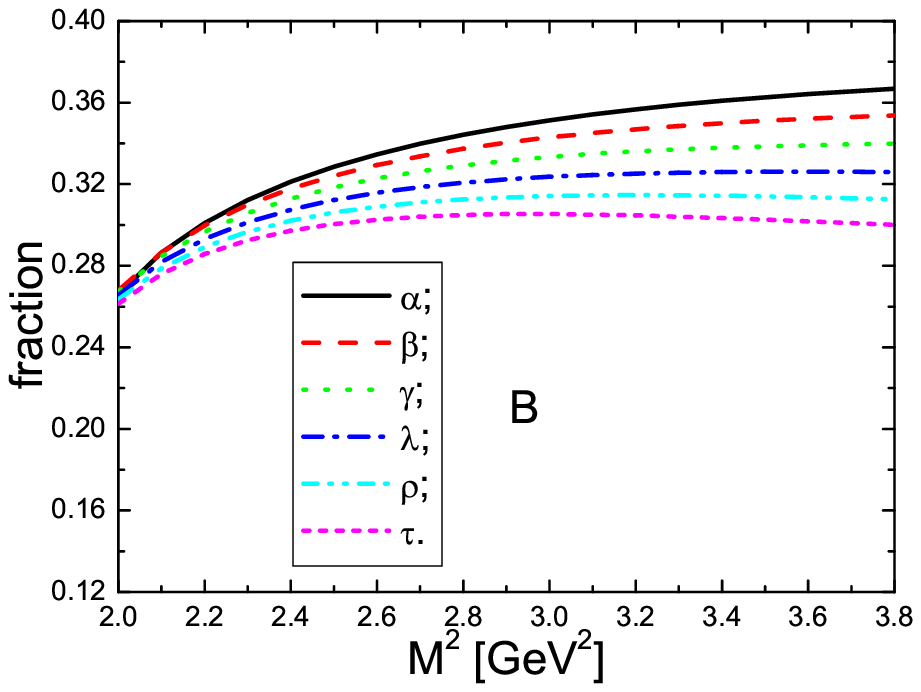}
 \includegraphics[totalheight=5cm,width=6cm]{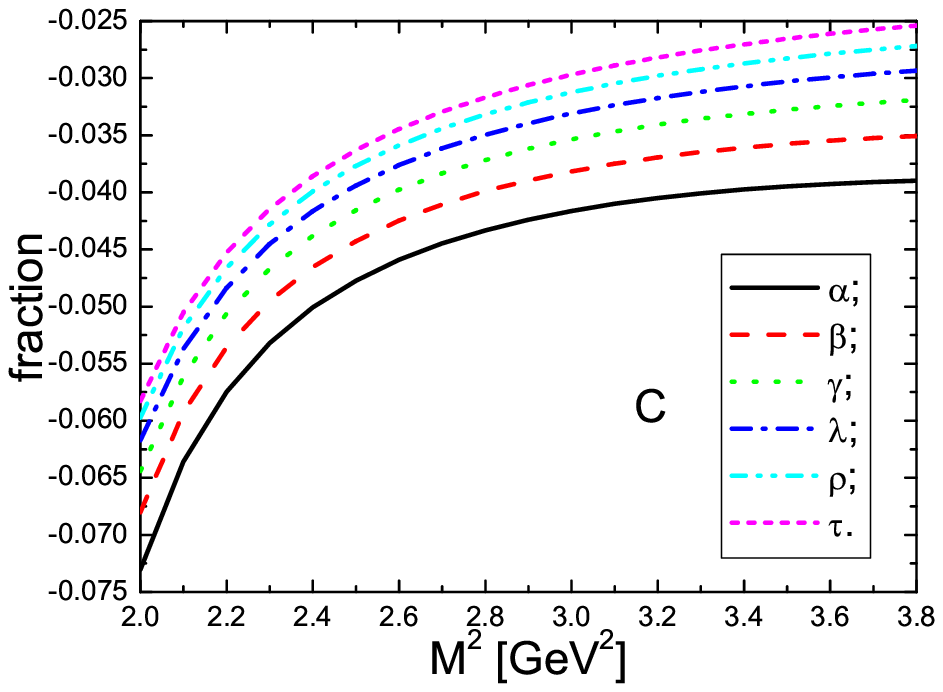}
 \includegraphics[totalheight=5cm,width=6cm]{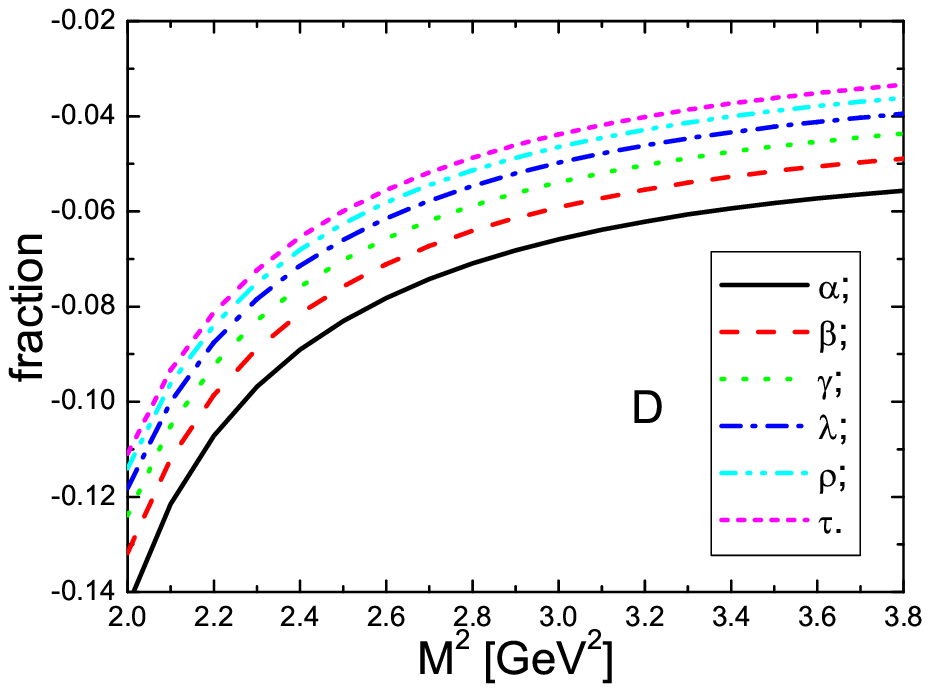}
 \includegraphics[totalheight=5cm,width=6cm]{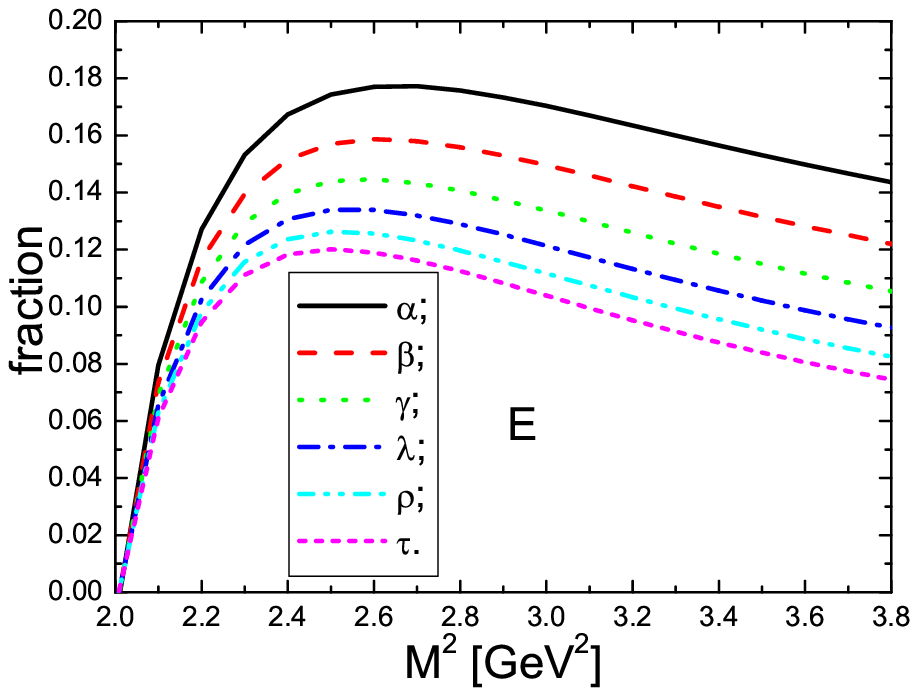}
 \includegraphics[totalheight=5cm,width=6cm]{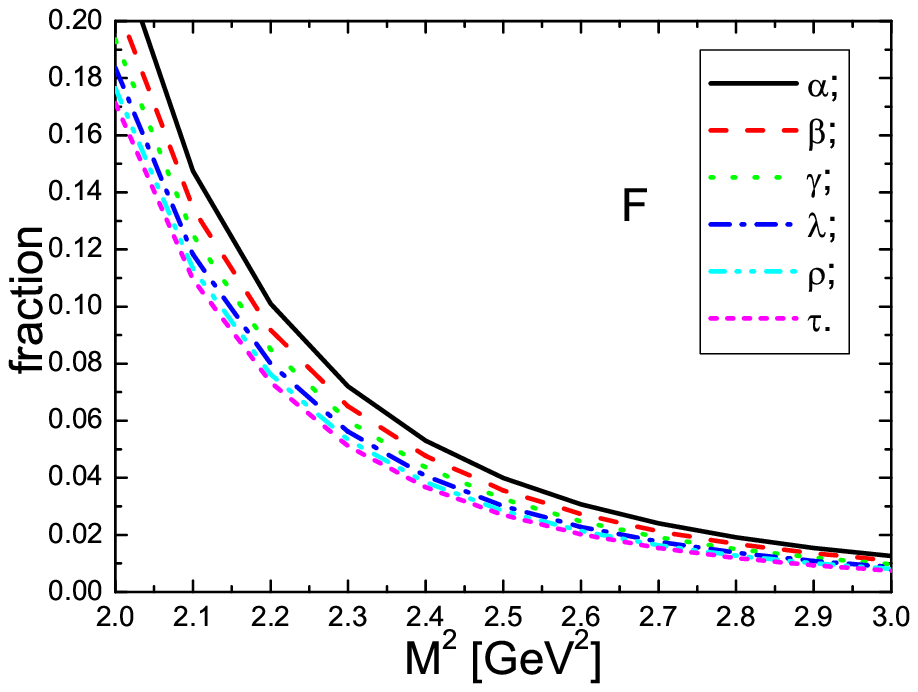}
    \caption{ The contributions from the different terms  with variation of the Borel parameter $M^2$ in the
    operator product expansion for the current $\bar{c}\gamma_\mu u \bar{d} \gamma^\mu c$. The $A$,
   $B$, $C$, $D$, $E$ and $F$ correspond to the contributions from
   the perturbative term,
$\langle \bar{s} s \rangle+\langle \bar{s}g_s\sigma G s \rangle$
term,  $\langle \frac{\alpha_s GG}{\pi} \rangle $ term, $\langle
\frac{\alpha_s GG}{\pi} \rangle $+$\langle \frac{\alpha_s GG}{\pi}
\rangle \left[\langle \bar{s} s \rangle +\langle \bar{s}g_s\sigma G
s \rangle+ \langle \bar{s}s \rangle^2\right]$ term, $\langle \bar{s}
s \rangle^2$+$\langle \bar{s} s \rangle\langle \bar{s}g_s\sigma G s
\rangle$ term and    $\langle \bar{s}g_s\sigma G s \rangle^2$ term,
respectively.     The notations
   $\alpha$, $\beta$, $\gamma$, $\lambda$, $\rho$ and $\tau$ correspond to the threshold
   parameters $s_0=21\,\rm{GeV}^2$,
   $22\,\rm{GeV}^2$, $23\,\rm{GeV}^2$, $24\,\rm{GeV}^2$, $25\,\rm{GeV}^2$ and $26\,\rm{GeV}^2$, respectively.
    Here we take the central values of the input parameters. }
\end{figure}

\begin{figure}
 \centering
 \includegraphics[totalheight=5cm,width=6cm]{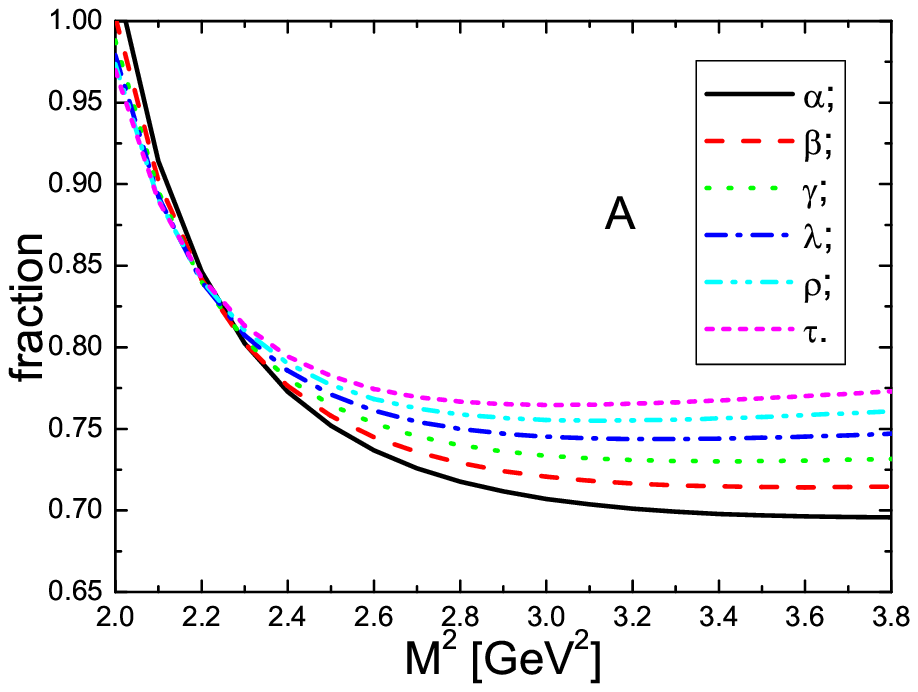}
 \includegraphics[totalheight=5cm,width=6cm]{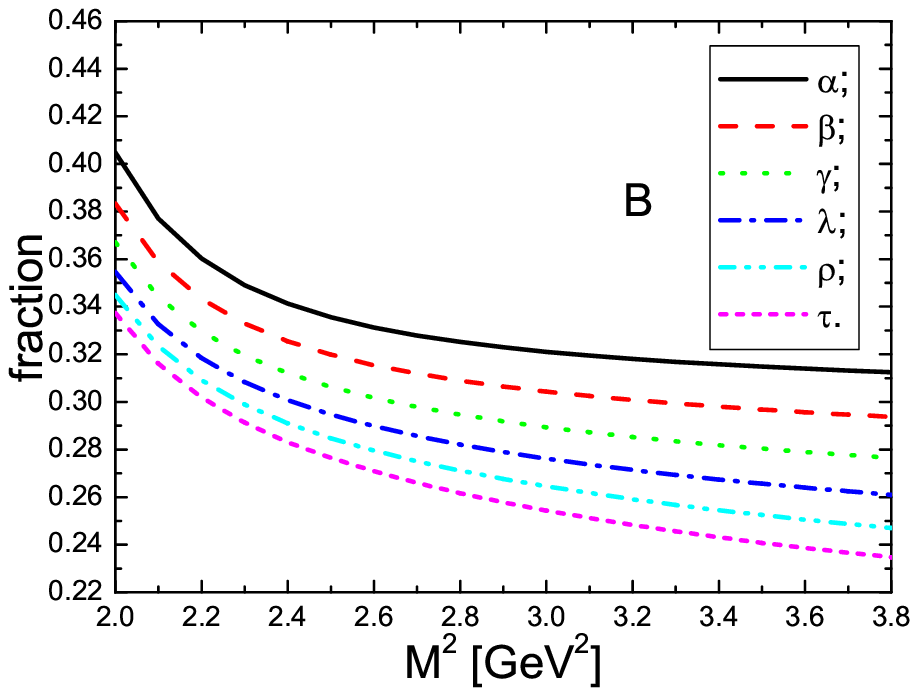}
 \includegraphics[totalheight=5cm,width=6cm]{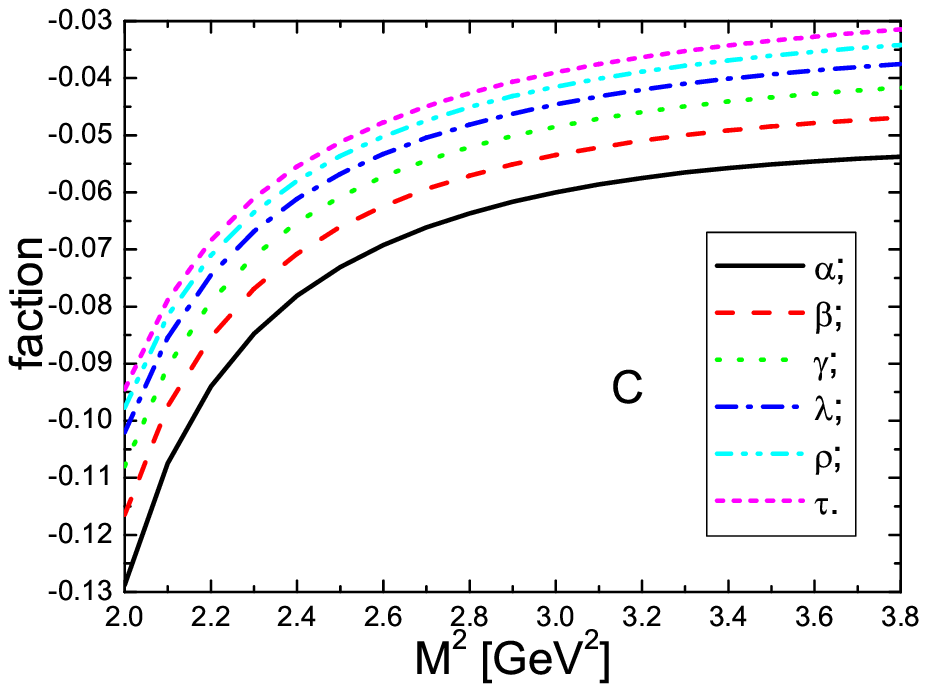}
 \includegraphics[totalheight=5cm,width=6cm]{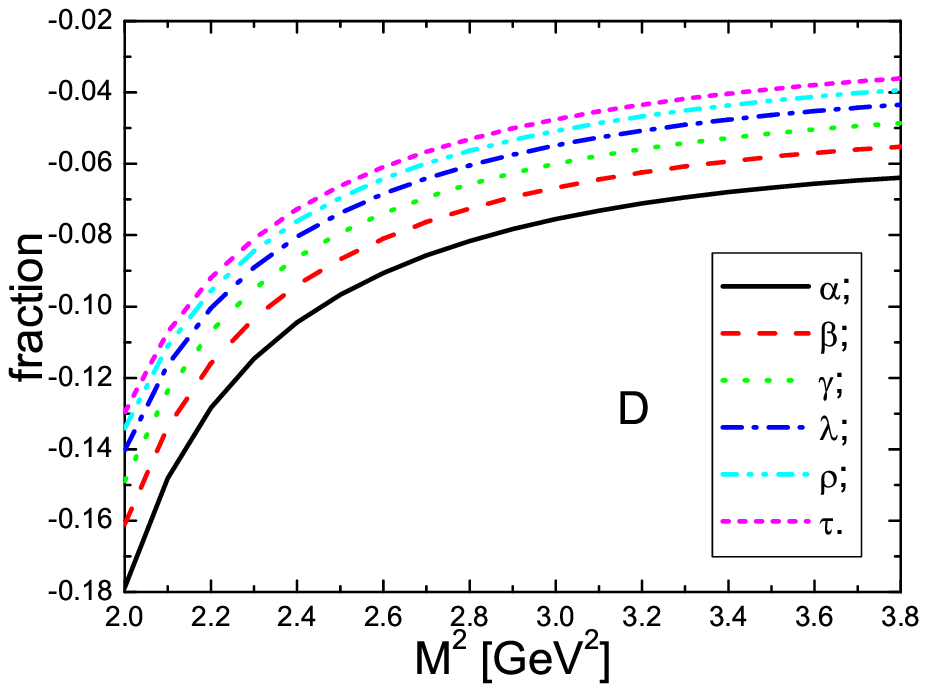}
 \includegraphics[totalheight=5cm,width=6cm]{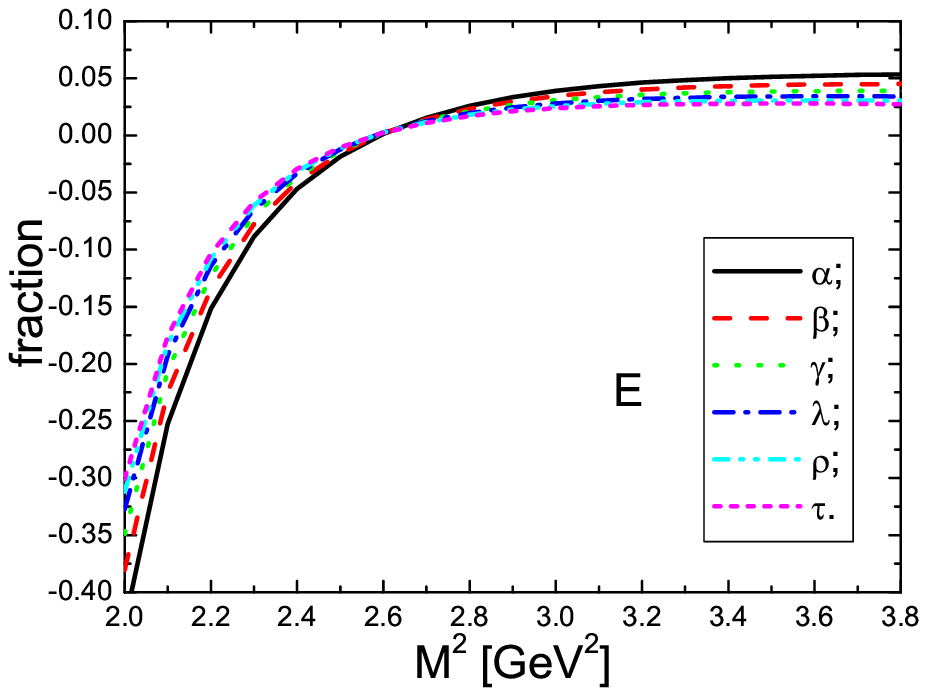}
 \includegraphics[totalheight=5cm,width=6cm]{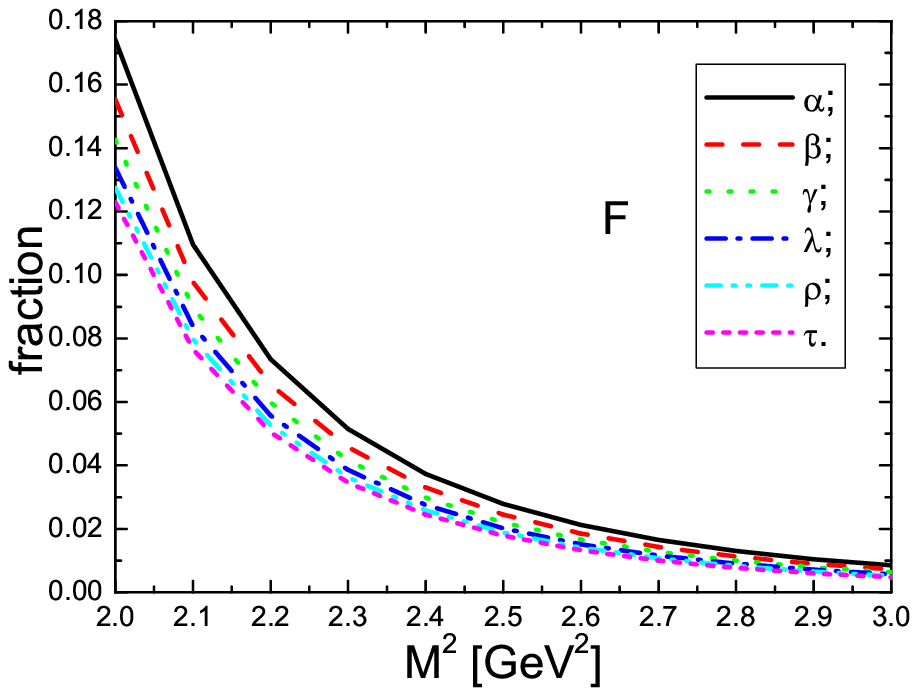}
    \caption{ The contributions from the different terms  with variation of the Borel parameter $M^2$ in the
    operator product expansion for the current $\bar{c}\gamma_\mu s \bar{s} \gamma^\mu c$. The $A$,
   $B$, $C$, $D$, $E$ and $F$ correspond to the contributions from
   the perturbative term,
$\langle \bar{s} s \rangle+\langle \bar{s}g_s\sigma G s \rangle$
term,  $\langle \frac{\alpha_s GG}{\pi} \rangle $ term, $\langle
\frac{\alpha_s GG}{\pi} \rangle $+$\langle \frac{\alpha_s GG}{\pi}
\rangle \left[\langle \bar{s} s \rangle +\langle \bar{s}g_s\sigma G
s \rangle+ \langle \bar{s}s \rangle^2\right]$ term, $\langle \bar{s}
s \rangle^2$+$\langle \bar{s} s \rangle\langle \bar{s}g_s\sigma G s
\rangle$ term and    $\langle \bar{s}g_s\sigma G s \rangle^2$ term,
respectively.     The notations
   $\alpha$, $\beta$, $\gamma$, $\lambda$, $\rho$ and $\tau$ correspond to the threshold
   parameters $s_0=21\,\rm{GeV}^2$,
   $22\,\rm{GeV}^2$, $23\,\rm{GeV}^2$, $24\,\rm{GeV}^2$, $25\,\rm{GeV}^2$ and $26\,\rm{GeV}^2$, respectively.
    Here we take the central values of the input parameters. }
\end{figure}

\begin{figure}
 \centering
 \includegraphics[totalheight=5cm,width=6cm]{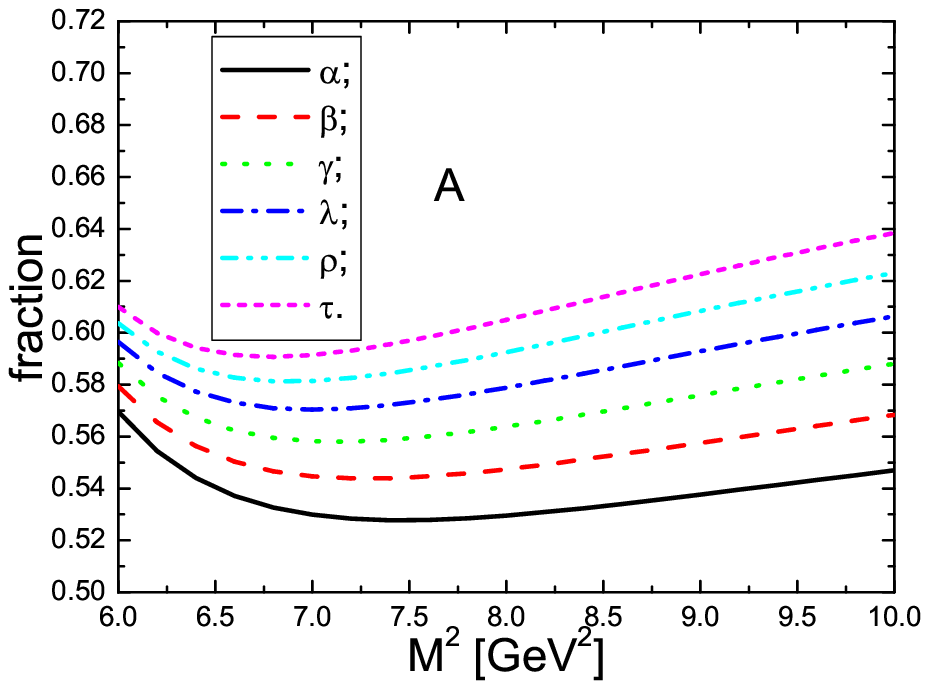}
 \includegraphics[totalheight=5cm,width=6cm]{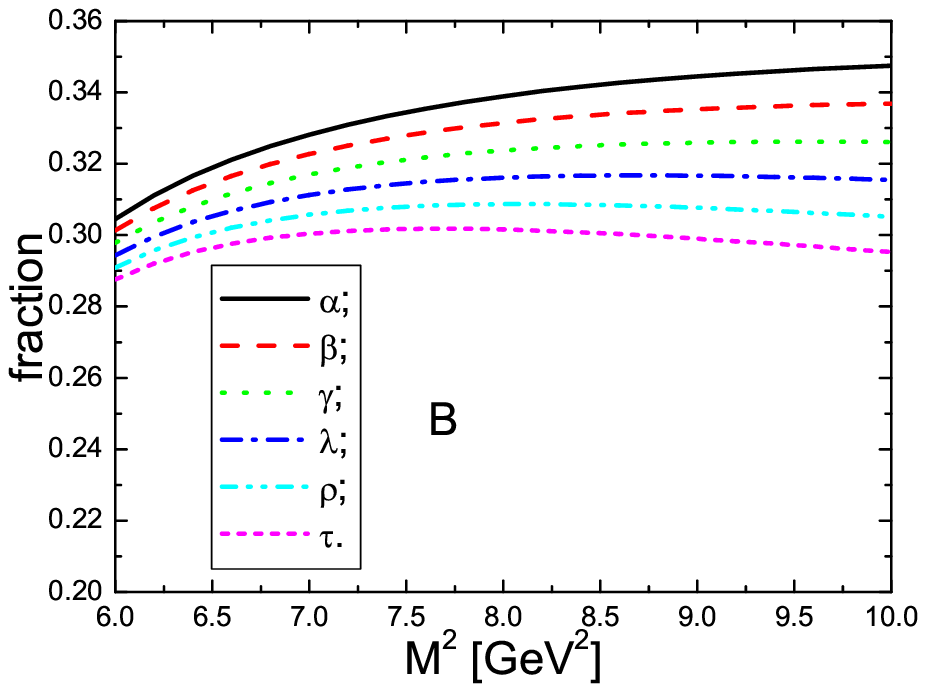}
 \includegraphics[totalheight=5cm,width=6cm]{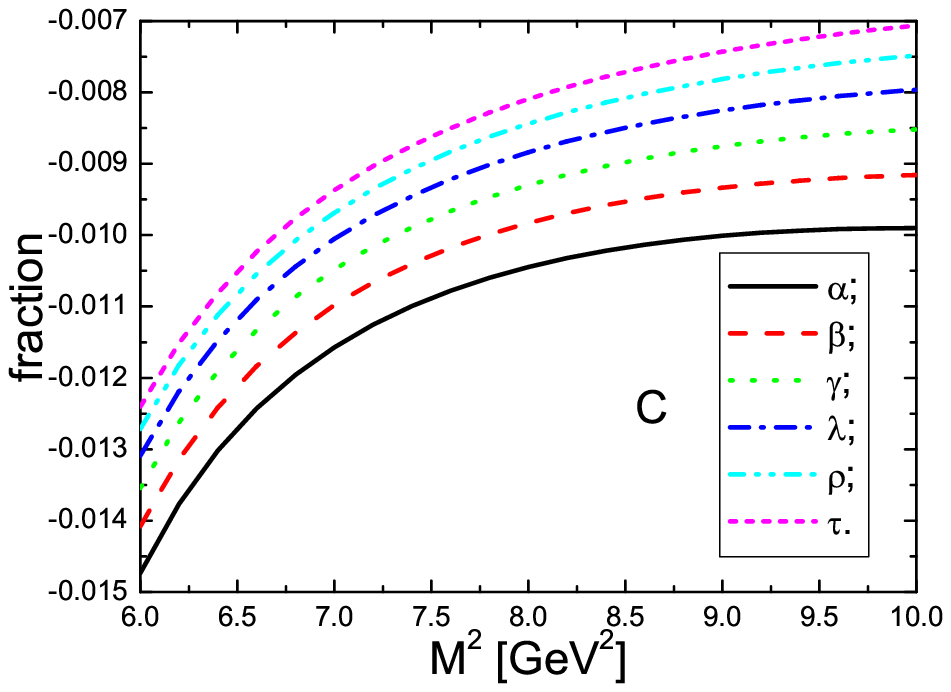}
 \includegraphics[totalheight=5cm,width=6cm]{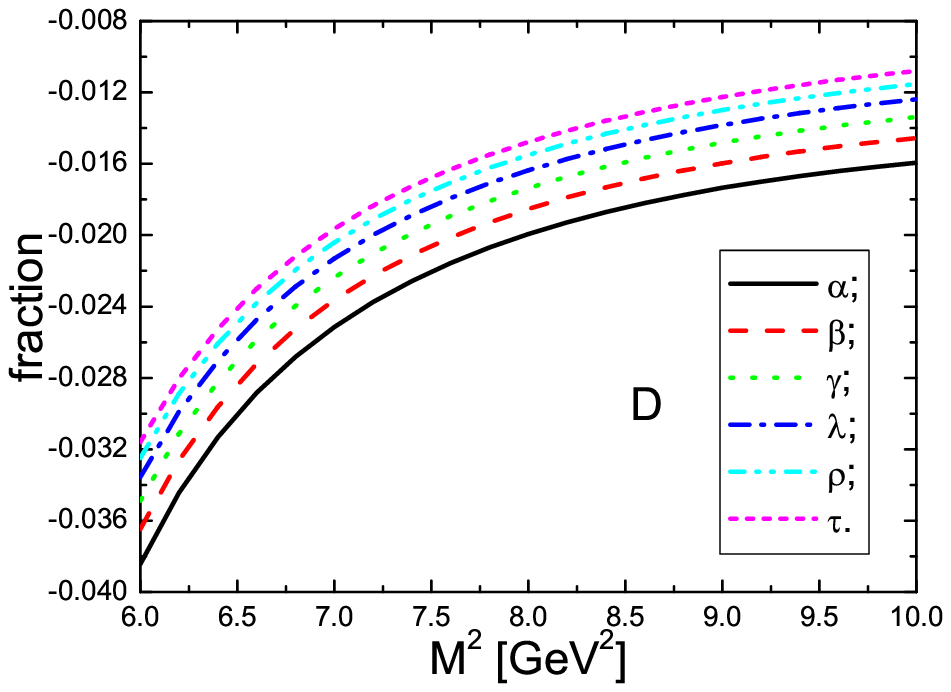}
 \includegraphics[totalheight=5cm,width=6cm]{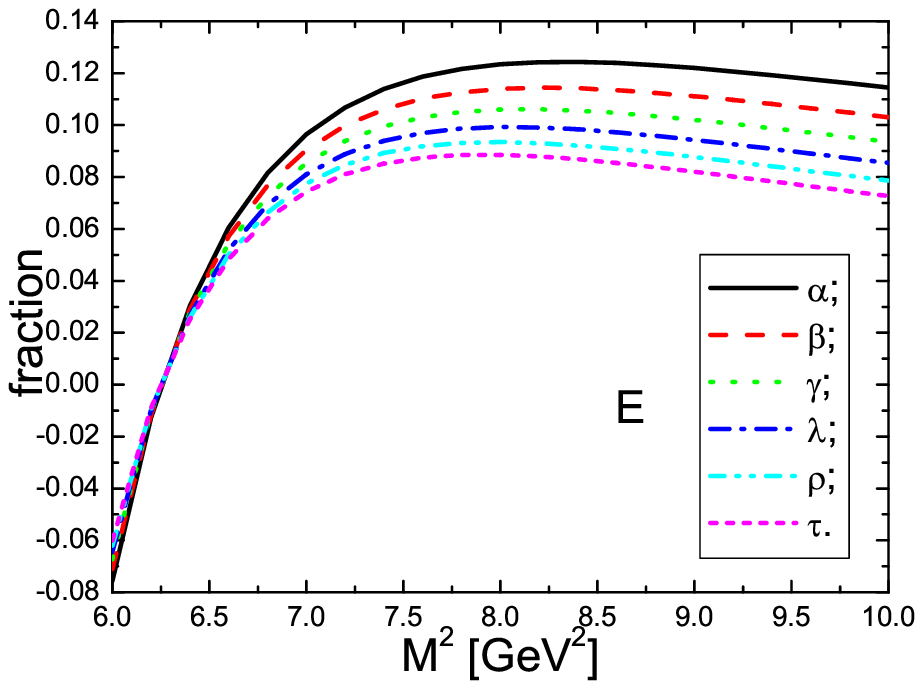}
 \includegraphics[totalheight=5cm,width=6cm]{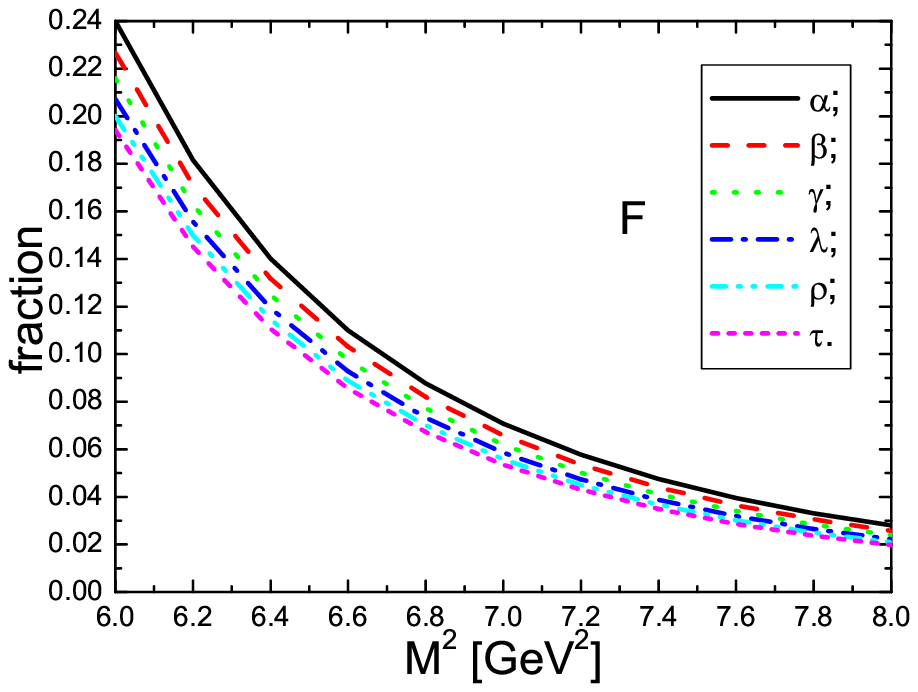}
    \caption{ The contributions from the different terms  with variation of the Borel parameter $M^2$ in the
    operator product expansion for the current $\bar{b}\gamma_\mu u \bar{d} \gamma^\mu b$. The $A$,
   $B$, $C$, $D$, $E$ and $F$ correspond to the contributions from
   the perturbative term,
$\langle \bar{s} s \rangle+\langle \bar{s}g_s\sigma G s \rangle$
term,  $\langle \frac{\alpha_s GG}{\pi} \rangle $ term, $\langle
\frac{\alpha_s GG}{\pi} \rangle $+$\langle \frac{\alpha_s GG}{\pi}
\rangle \left[\langle \bar{s} s \rangle +\langle \bar{s}g_s\sigma G
s \rangle+ \langle \bar{s}s \rangle^2\right]$ term, $\langle \bar{s}
s \rangle^2$+$\langle \bar{s} s \rangle\langle \bar{s}g_s\sigma G s
\rangle$ term and    $\langle \bar{s}g_s\sigma G s \rangle^2$ term,
respectively.     The notations
   $\alpha$, $\beta$, $\gamma$, $\lambda$, $\rho$ and $\tau$ correspond to the threshold
   parameters $s_0=132\,\rm{GeV}^2$,
   $134\,\rm{GeV}^2$, $136\,\rm{GeV}^2$, $138\,\rm{GeV}^2$, $140\,\rm{GeV}^2$ and $142\,\rm{GeV}^2$, respectively.
    Here we take the central values of the input parameters. }
\end{figure}

\begin{figure}
 \centering
 \includegraphics[totalheight=5cm,width=6cm]{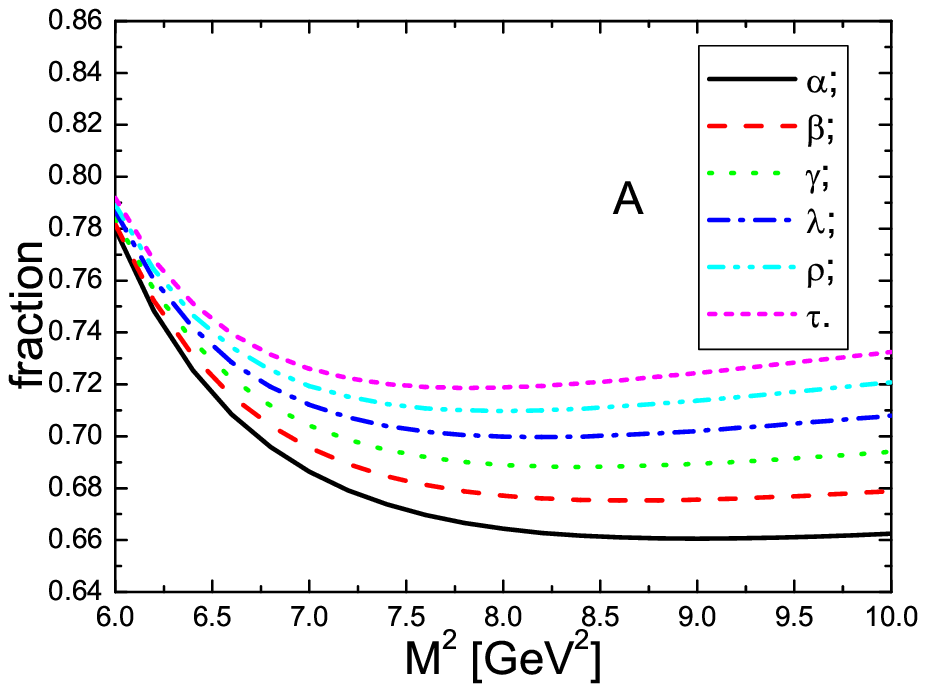}
 \includegraphics[totalheight=5cm,width=6cm]{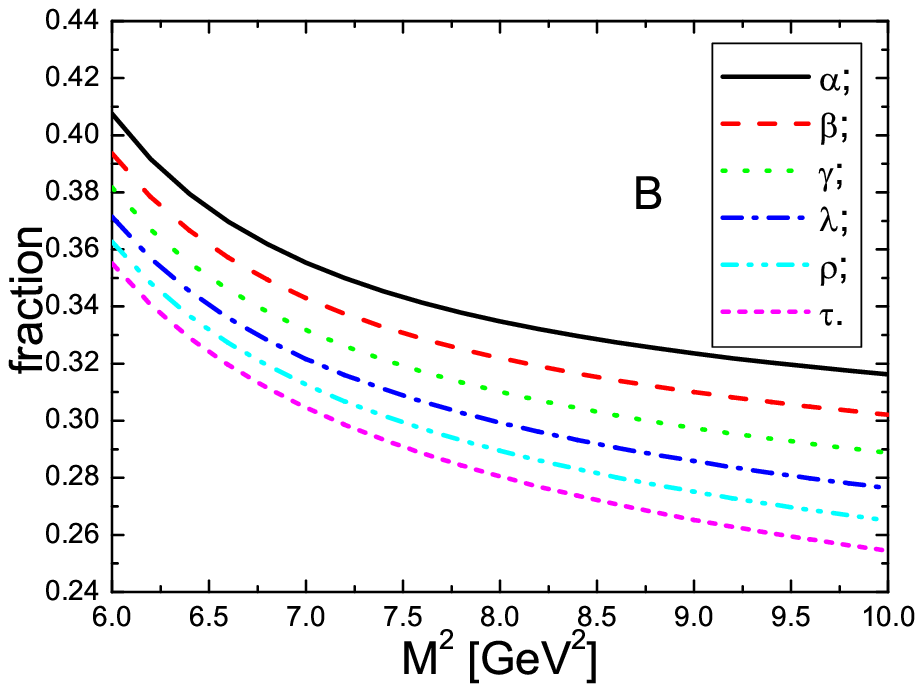}
 \includegraphics[totalheight=5cm,width=6cm]{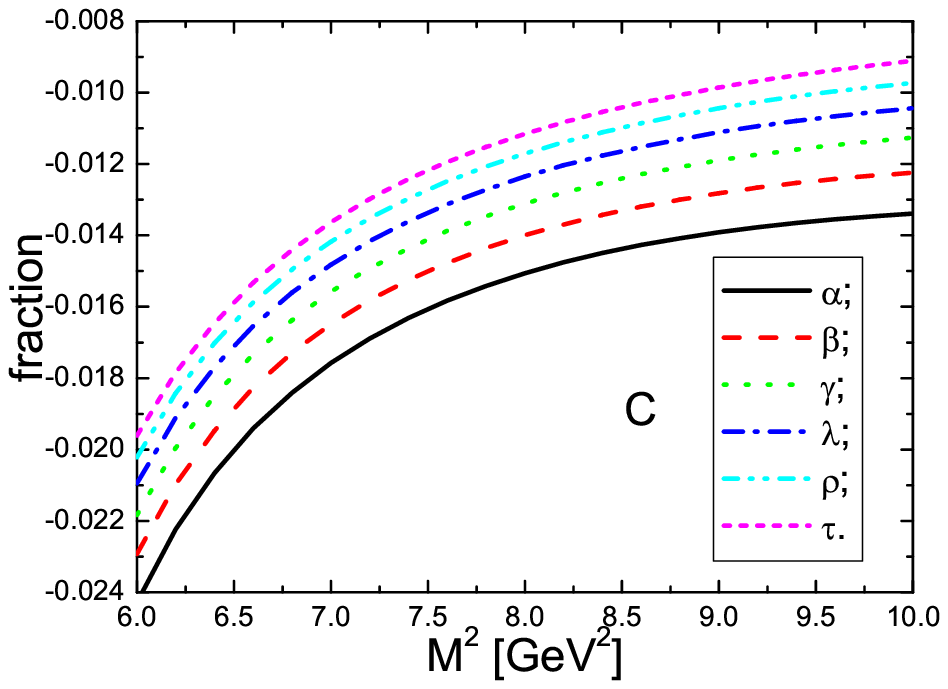}
 \includegraphics[totalheight=5cm,width=6cm]{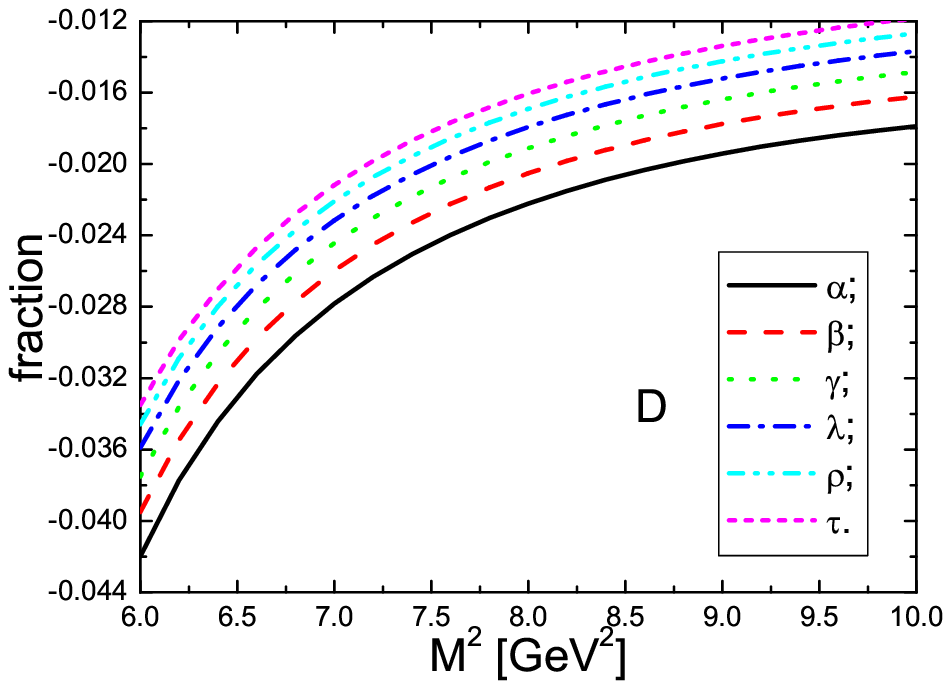}
 \includegraphics[totalheight=5cm,width=6cm]{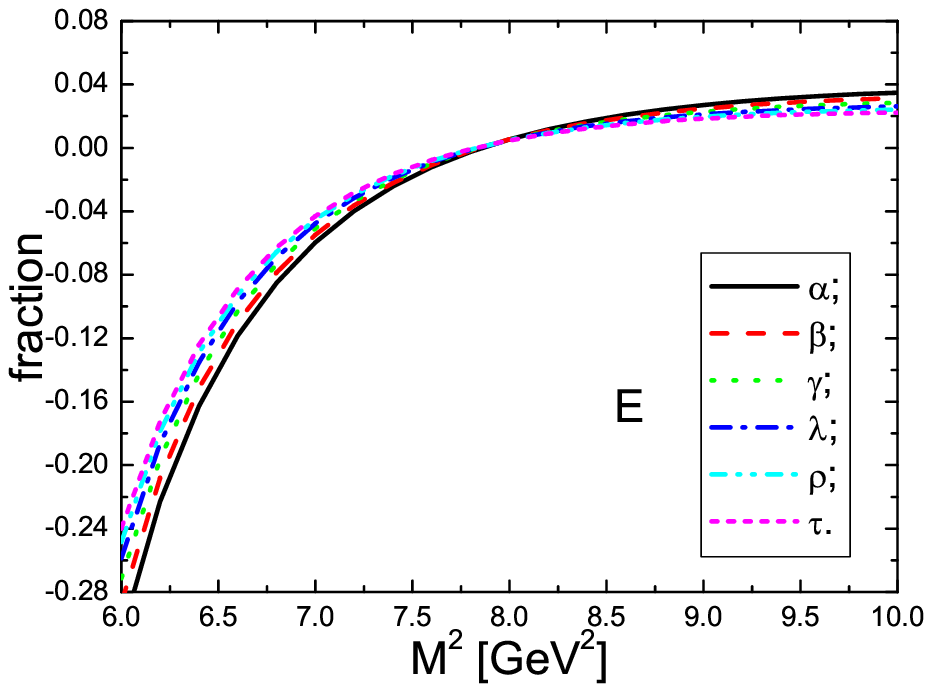}
 \includegraphics[totalheight=5cm,width=6cm]{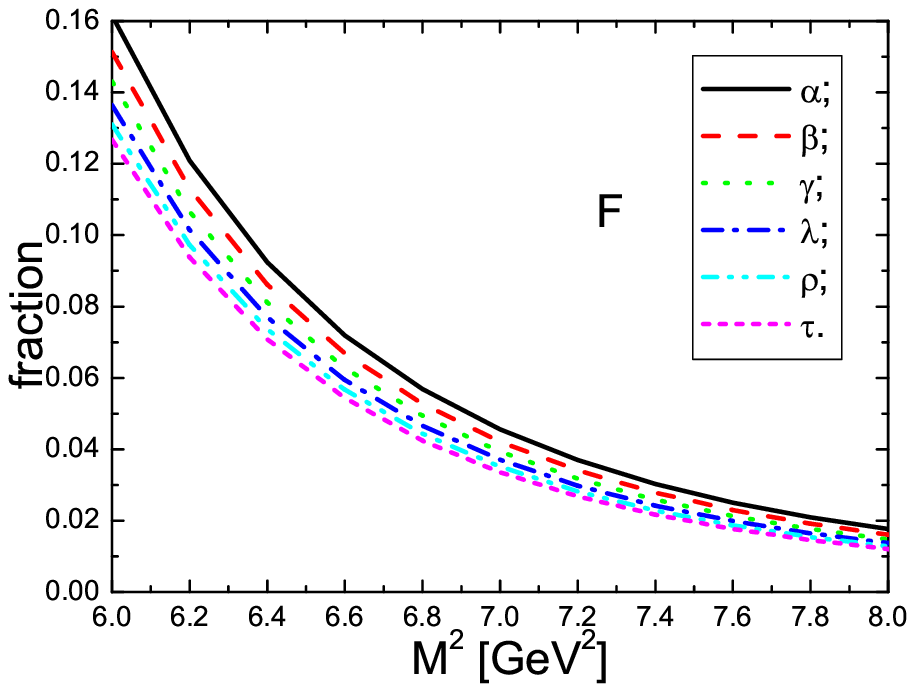}
    \caption{ The contributions from the different terms  with variation of the Borel parameter $M^2$ in the
    operator product expansion for the current $\bar{b}\gamma_\mu s \bar{s} \gamma^\mu b$. The $A$,
   $B$, $C$, $D$, $E$ and $F$ correspond to the contributions from
   the perturbative term,
$\langle \bar{s} s \rangle+\langle \bar{s}g_s\sigma G s \rangle$
term,  $\langle \frac{\alpha_s GG}{\pi} \rangle $ term, $\langle
\frac{\alpha_s GG}{\pi} \rangle $+$\langle \frac{\alpha_s GG}{\pi}
\rangle \left[\langle \bar{s} s \rangle +\langle \bar{s}g_s\sigma G
s \rangle+ \langle \bar{s}s \rangle^2\right]$ term, $\langle \bar{s}
s \rangle^2$+$\langle \bar{s} s \rangle\langle \bar{s}g_s\sigma G s
\rangle$ term and    $\langle \bar{s}g_s\sigma G s \rangle^2$ term,
respectively.     The notations
   $\alpha$, $\beta$, $\gamma$, $\lambda$, $\rho$ and $\tau$ correspond to the threshold
   parameters $s_0=132\,\rm{GeV}^2$,
   $134\,\rm{GeV}^2$, $136\,\rm{GeV}^2$, $138\,\rm{GeV}^2$, $140\,\rm{GeV}^2$ and $142\,\rm{GeV}^2$, respectively.
    Here we take the central values of the input parameters. }
\end{figure}

The contributions from the different terms   in the operator product
expansion are shown in Figs.1-4, where (and thereafter) we
 use the $\langle\bar{s}s\rangle$ to denote the quark condensates
$\langle\bar{q}q\rangle$, $\langle\bar{s}s\rangle$ and the
$\langle\bar{s}g_s \sigma Gs\rangle$ to denote the mixed condensates
$\langle\bar{q}g_s \sigma Gq\rangle$, $\langle\bar{s}g_s \sigma
Gs\rangle$. From the figures, we can see that the contributions from
 different terms in the operator product
expansion change  quickly with variation of the Borel parameter at
the values $M^2\leq 2.6\,\rm{GeV}^2$ and $M^2\leq 7.0\,\rm{GeV}^2$
in the hidden charm  and hidden bottom channels respectively, such
an unstable  behavior  can not lead to stable sum rules, our
numerical results confirm this conjecture, see Figs.6-7.

 The dominant  contributions come from the perturbative term and the $\langle
\bar{s}s\rangle+ \langle \bar{s}g_s \sigma Gs\rangle $ term; and the
interpolating currents contain more $s$ quarks have better
convergent behavior. The contribution from the terms involving the
gluon condensate $\langle \frac{\alpha_s GG}{\pi}\rangle$ are very
small, the gluon condensate plays a minor important role. The vacuum
condensates of the high dimension $\langle
\bar{s}s\rangle^2$+$\langle \bar{s}s\rangle\langle \bar{s} g_s
\sigma G s\rangle$ serve as a criterion for choosing the Borel
parameter $M^2$ and threshold parameter $s_0$.

At the values $M_{min}^2\geq 2.6\,\rm{GeV}^2$ and $s_0\geq
23\,\rm{GeV}^2$, the contributions from the high dimensional
condensates  $\langle \bar{s}s\rangle^2+\langle \bar{s}s\rangle
\langle \bar{s}g_s \sigma Gs\rangle $  are less than  $15\%$ ($4\%$)
in the $\bar{c}\gamma_\mu u \bar{d} \gamma^\mu c$
($\bar{c}\gamma_\mu s \bar{s} \gamma^\mu c$) channel; the
contributions from the vacuum condensate of the highest dimension
$\langle\bar{s}g_s \sigma Gs\rangle^2$ are less than  $3\%$ in all
the hidden charm channels, we expect the operator product expansion
is convergent in the hidden charm channels. At the values
$M_{min}^2\geq 7.0\,\rm{GeV}^2$ and $s_0\geq 136\,\rm{GeV}^2$, the
contributions from the high dimensional condensates  $\langle
\bar{s}s\rangle^2+\langle \bar{s}s\rangle \langle \bar{s}g_s \sigma
Gs\rangle $  are less than $11\%$ ($5\%$) in the $\bar{b}\gamma_\mu
u \bar{d} \gamma^\mu b$ ($\bar{b}\gamma_\mu s \bar{s} \gamma^\mu b$)
channel; the contributions from the vacuum condensate of the highest
dimension $\langle\bar{s}g_s \sigma Gs\rangle^2$ are less than $7\%$
in all the hidden bottom channels, we expect the operator product
expansion is convergent in the hidden bottom channels.

 In this article, we take the uniform Borel parameter
$M^2_{min}$, i.e. $M^2_{min}\geq 2.6 \, \rm{GeV}^2$ and
$M^2_{min}\geq 7.0 \, \rm{GeV}^2$ in the hidden charm and hidden
bottom channels, respectively.

In Fig.5, we show the  contributions from the pole terms with
variation of the Borel parameter and the threshold parameter. The
pole contributions are larger than  $51\%$ ($55\%$) at the value
$M_{max}^2 \leq 3.0 \, \rm{GeV}^2 $ and $s_0\geq 23\,\rm{GeV}^2$ ($
24\,\rm{GeV}^2$) in the $\bar{c}\gamma_\mu u \bar{d} \gamma^\mu c$
($\bar{c}\gamma_\mu s \bar{s} \gamma^\mu c$) channel, and larger
than  $52\%$ ($55\%$) at the value $M_{max}^2 \leq 8.0 \, \rm{GeV}^2
$ and $s_0\geq 136\,\rm{GeV}^2$ ($ 138\,\rm{GeV}^2$) in the
$\bar{b}\gamma_\mu u \bar{d} \gamma^\mu b$ ($\bar{b}\gamma_\mu s
\bar{s} \gamma^\mu b$) channel. Again we take the uniform Borel
parameter $M^2_{max}$, i.e. $M^2_{max}\leq 3.0 \, \rm{GeV}^2$ and
$M^2_{max}\leq 8.0 \, \rm{GeV}^2$ in the hidden charm  and hidden
bottom channels, respectively.

In this article, the threshold parameters are taken as
$s_0=(24\pm1)\,\rm{GeV}^2$, $(25\pm1)\,\rm{GeV}^2$,
$(138\pm2)\,\rm{GeV}^2$,  and $(140\pm2)\,\rm{GeV}^2$ in the
$\bar{c}\gamma_\mu u \bar{d} \gamma^\mu c$,
   $\bar{c}\gamma_\mu s \bar{s} \gamma^\mu c$, $\bar{b}\gamma_\mu u \bar{d} \gamma^\mu b$,
    and $\bar{b}\gamma_\mu s \bar{s} \gamma^\mu b$ channels, respectively;
   the Borel parameters are taken as $M^2=(2.6-3.0)\,\rm{GeV}^2$ and
   $(7.0-8.0)\,\rm{GeV}^2$ in the
hidden charm  and hidden bottom channels, respectively.
      In those regions, the two criteria of the QCD sum rules
are full satisfied  \cite{SVZ79,Reinders85}.

\begin{figure}
\centering
\includegraphics[totalheight=5cm,width=6cm]{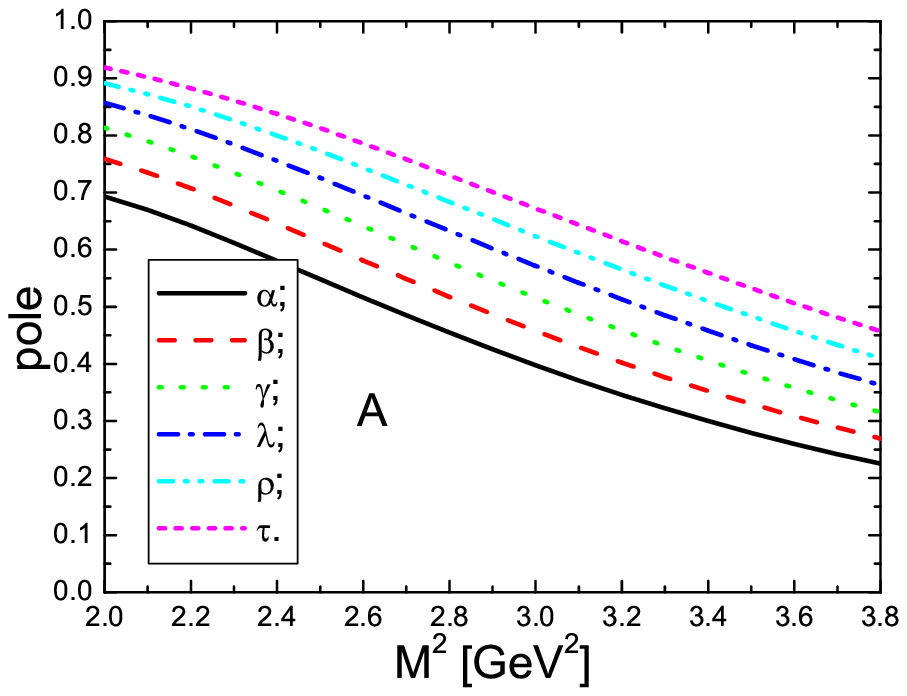}
\includegraphics[totalheight=5cm,width=6cm]{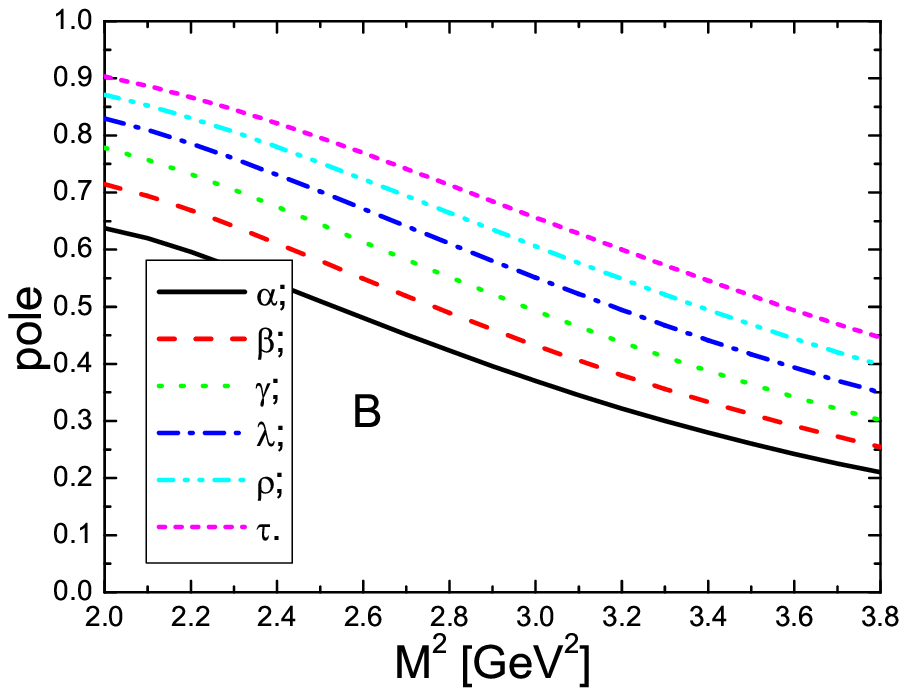}
\includegraphics[totalheight=5cm,width=6cm]{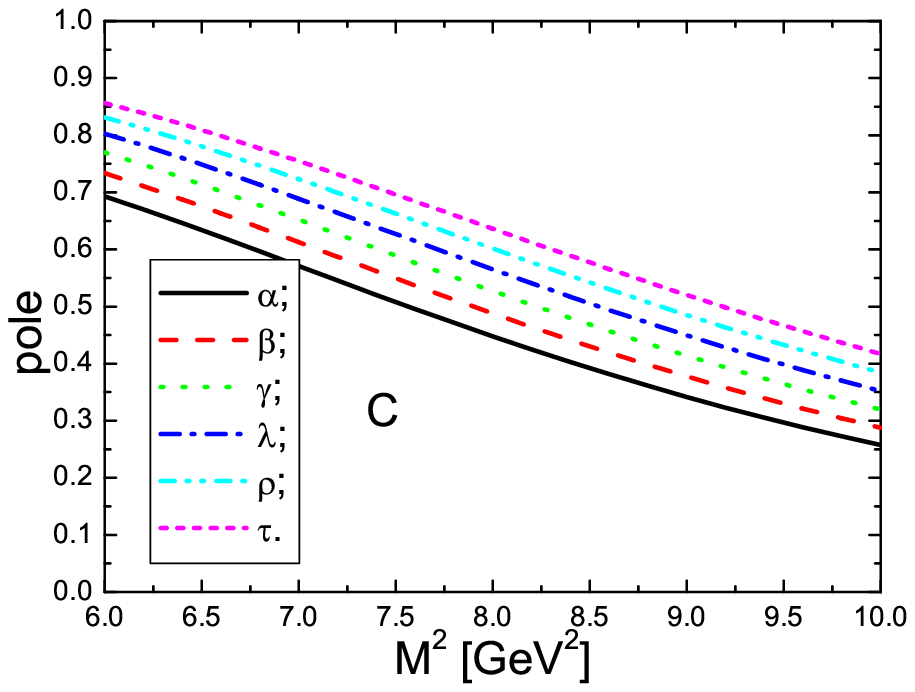}
\includegraphics[totalheight=5cm,width=6cm]{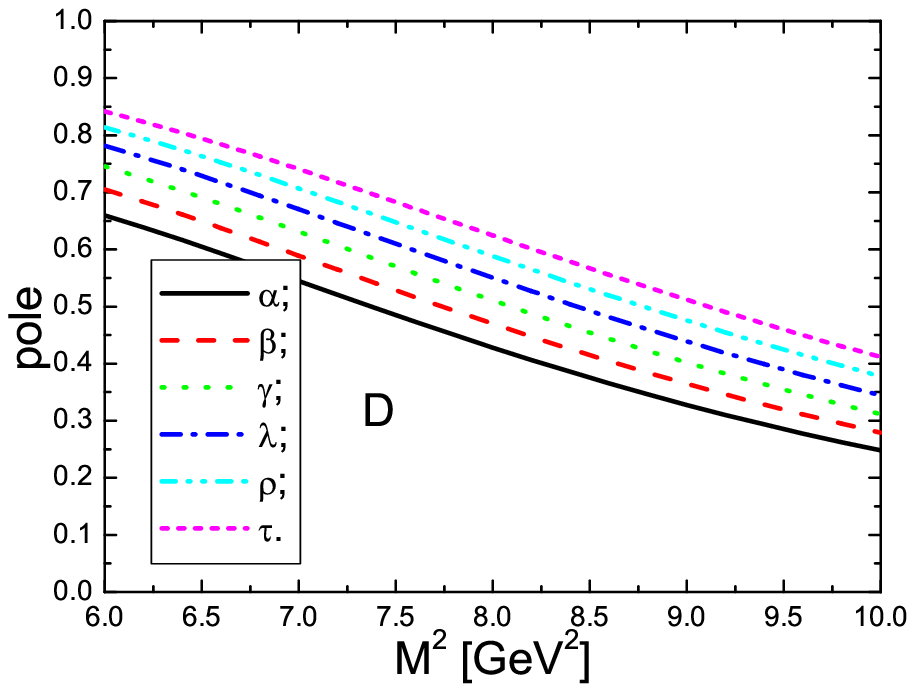}
  \caption{ The contributions from the pole terms with variation of the Borel parameter $M^2$. The $A$, $B$, $C$
   and $D$ denote the $\bar{c}\gamma_\mu u \bar{d} \gamma^\mu c$,
   $\bar{c}\gamma_\mu s \bar{s} \gamma^\mu c$, $\bar{b}\gamma_\mu u \bar{d} \gamma^\mu b$,
   and $\bar{b}\gamma_\mu s \bar{s} \gamma^\mu b$  channels, respectively.  In the hidden charm channels, the notations
   $\alpha$, $\beta$, $\gamma$, $\lambda$, $\rho$ and $\tau$  correspond to the threshold
   parameters $s_0=21\,\rm{GeV}^2$,
   $22\,\rm{GeV}^2$, $23\,\rm{GeV}^2$, $24\,\rm{GeV}^2$, $25\,\rm{GeV}^2$ and $26\,\rm{GeV}^2$ respectively
   ;  while in the hidden bottom channels they correspond to
    the threshold
   parameters  $s_0=132\,\rm{GeV}^2$,
   $134\,\rm{GeV}^2$, $136\,\rm{GeV}^2$, $138\,\rm{GeV}^2$, $140\,\rm{GeV}^2$ and $142\,\rm{GeV}^2$ respectively.   }
\end{figure}

Taking into account all uncertainties of the input parameters,
finally we obtain the values of the masses and pole residues  of
 the scalar molecular states   $Y$, which are  shown in Figs.6-7 and Tables 1-2.

 From Tables 1-2, we can see that the uncertainties of the masses
$M_Y$ are rather small (about $4\%$ in the hidden charm channels and
$2\%$ in the hidden bottom channels) while the uncertainties of the
pole residues $\lambda_{Y}$ are rather large (about $(18-22)\%$).
The uncertainties of the input parameters ($\langle \bar{q}q
\rangle$, $\langle \bar{s}s \rangle$, $\langle \bar{s}g_s\sigma G s
\rangle$, $\langle \bar{q}g_s\sigma G q \rangle$,
 $m_s$,  $m_c$ and $m_b$) vary in the range
$(2-25)\%$, the uncertainties of the pole  residues $\lambda_{Y}$
are reasonable. We obtain the  squared masses   $M_Y^2$ through a
fraction, the uncertainties in the numerator and denominator which
originate from a given input parameter (for example, $\langle
\bar{s}s \rangle$, $\langle \bar{s}g_s\sigma G s \rangle$) cancel
out with each other, and result in small net uncertainty.

At the energy scale $\mu=1\, \rm{GeV}$, $\frac{\alpha_s}{\pi}\approx
0.19$ \cite{Davier2006}, if the perturbative $\mathcal
{O}(\alpha_s)$ corrections to the perturbative term are companied
with large numerical factors, $1+\xi(s,m_Q)\frac{\alpha_s}{\pi}$,
 the contributions may be large. For example, we can make a crude estimation by
multiplying the perturbative term  with a numerical factor, say
$1+\xi(s,m_c)\frac{\alpha_s}{\pi}=2$, in the hidden charm channel,
the mass $M_{D_s^\ast {\bar D}_s^\ast}$ decreases slightly, while
the  pole residue $\lambda_{D_s^\ast {\bar D}_s^\ast}$ increases
remarkably. From Figs.1-4, we can see that the main contributions
come from the perturbative terms, the large corrections in the
numerator and denominator cancel out with each other. In fact, the
$\xi(s,m_Q)$ are complicated functions of the energy $s$ and the
mass $m_Q$, such a crude estimation maybe underestimate the
$\mathcal {O}(\alpha_s)$ corrections, the uncertainties originate
from the $\mathcal {O}(\alpha_s)$ corrections maybe larger.

In this article, we also neglect  the contributions from the
perturbative  corrections $\mathcal {O}(\alpha_s^n)$.  Those
perturbative  corrections can be taken into account in the leading
logarithmic
 approximations through  anomalous dimension factors. After the Borel transform, the effects of those
 corrections are  to multiply each term on the operator product
 expansion side by the factor,
 \begin{eqnarray}
 \left[ \frac{\alpha_s(M^2)}{\alpha_s(\mu^2)}\right]^{2\Gamma_{J/\eta}-\Gamma_{\mathcal
 {O}_n}} \, ,
 \end{eqnarray}
 where the $\Gamma_{J/\eta}$ is the anomalous dimension of the scalar
 interpolating current $J/\eta(x)$, the $\Gamma_{\mathcal {O}_n}$ is the anomalous dimension of
 the local operator $\mathcal {O}_n(0)$ in the operator product
 expansion,
 \begin{eqnarray}
 T\left\{J/\eta(x)J/\eta^{\dagger}(0)\right\}&=&C_n(x) {O}_n(0) \, ,
 \end{eqnarray}
here the $C_n(x)$ is the corresponding Wilson coefficient.

We carry out the operator product expansion at a special energy
scale, say $\mu=1\,\rm{GeV}$, and  can  not smear the scale
dependence by evolving the operator product    expansion side to the
energy scale $M$ through Eq.(14) as the anomalous dimension of the
scalar current $J/\eta(x)$ is unknown.
 Furthermore, the anomalous
dimensions of the high dimensional local operators have not been
calculated yet, and their values are poorly known.  In this article,
we set the factor $\left[
\frac{\alpha_s(M^2)}{\alpha_s(\mu^2)}\right]^{2\Gamma_{J/\eta}-\Gamma_{\mathcal
{O}_n}}\approx1$, such an approximation maybe result in some scale
dependence  and  weaken the prediction ability; further studies are
stilled needed.

The central value of the present prediction $M_{D_s^\ast {\bar
D}_s^\ast}=(4.48\pm0.17)\,\rm{GeV}$ is slightly larger than our
previous calculation $M_{D_s^\ast {\bar D}_s^\ast}=(4.43\pm0.16)
\rm{GeV}$ \cite{Wang0903}. In the present work, we take a slightly
larger threshold parameter $s_0=(25\pm1)\,\rm{GeV}^2$ rather than
$s_0=(24\pm1)\,\rm{GeV}^2$ to take into account the $SU(3)$ breading
effects and enhance the contribution from the pole term.  From Table
1, we can see that the central values of the possible scalar
molecular states are about $(250-500)\,\rm{MeV}$ above     the
corresponding $D^\ast -{\bar D}^\ast$,  $D_s^\ast -{\bar D}_s^\ast$,
$B^\ast- {\bar B}^\ast$, $B_s^\ast -{\bar B}_s^\ast$ thresholds
respectively \cite{PDG},  the $D^\ast {\bar D}^\ast$, $D_s^\ast
{\bar D}_s^\ast$, $B^\ast {\bar B}^\ast$, $B_s^\ast {\bar B}_s^\ast$
are probably virtual states. In the constituent quark models, the
energy gap between the ground state and the first radial excited
state is about $500\,\rm{MeV}$. The central values listed in Table 1
are below the corresponding thresholds of the
 first radial excited meson pairs. The scalar $D^\ast {\bar D}^\ast$,
$D_s^\ast {\bar D}_s^\ast$, $B^\ast {\bar B}^\ast$, $B_s^\ast {\bar
B}_s^\ast$ molecular states maybe not exist, while  the scalar
${D'}^\ast {\bar {D'}}^\ast$, ${D'}_s^\ast {\bar {D'}}_s^\ast$,
${B'}^\ast {\bar {B'}}^\ast$ and ${B'}_s^\ast {\bar {B'}}_s^\ast$
molecular states maybe exist.

In Refs.\cite{NielsenMolecule,Wang0903}, the same  current
$\bar{c}(x)\gamma_\mu s(x) \bar{s}(x)\gamma^\mu c(x)$  is used to
interpolate the narrow structure $Y(4140)$, however, the conclusions
are quite different. The discrepancy mainly originates from the high
dimensional vacuum condensates,  the vacuum condensates of
dimension-9,10 and the gluon involved vacuum condensates of
dimension larger than 4 are neglected in Ref.\cite{NielsenMolecule}.
Those condensates are counted as $1$, $\mathcal
{O}(\frac{m_c^2}{M^2})$, $\mathcal {O}(\frac{m_c^4}{M^4})$,
$\mathcal {O}(\frac{m_c^6}{M^6})$ respectively, and the
corresponding  contributions are greatly enhanced  at small $M^2$,
and result in rather  bad  convergent behavior in the operator
product expansion, we have to choose larger  Borel parameter $M^2$,
one can consult  the contributions from the $\langle \bar{s}g_s
\sigma G s\rangle^2$ term in Figs.1-F.2-F,3-F,4-F for example. If we
neglect the terms concerning those high dimensional vacuum
condensates  and choose the input parameters (especially the value
of the $m_c$) as Ref.\cite{NielsenMolecule}, the experimental data
can be reproduced. As a byproduct, we can see that the scale
dependence of the QCD sum rules only weakens  the prediction ability
mildly. However, we insist on taking into account the high
dimensional  vacuum condensates, as the interpolating current
consists  of  a light quark-antiquark pair and a heavy
quark-antiquark pair, one of the highest dimensional vacuum
condensates is $\langle \bar{s}s\rangle^2\times \langle
\frac{\alpha_s GG}{\pi}\rangle$.

The $c$-quark mass appearing in the perturbative terms (see e.g.
Eq.(18)) is usually taken to be the pole mass in the QCD sum rules,
while the choice of the $m_c$ in the leading-order coefficients of
the higher-dimensional terms is arbitrary \cite{Kho9801}.  The
$\overline{MS}$ mass $m_c(m_c^2)$ relates with the pole mass
$\hat{m}$ through the relation
\begin{eqnarray}
m_c(m_c^2) &=&\hat{m}\left[1+\frac{C_F
\alpha_s(m_c^2)}{\pi}+(K-2C_F)\left(\frac{\alpha_s}{\pi}\right)^2+\cdots\right]^{-1}\,
,
\end{eqnarray}
where $K$ depends on the flavor number $n_f$. In this article, we
take the approximation $m_c\approx\hat{m}$ without the $\alpha_s$
corrections for consistency. The value listed in the Particle Data
Group is $m_c(m_c^2)=1.27^{+0.07}_{-0.11} \, \rm{GeV}$ \cite{PDG},
it is reasonable to take the value
$m_c=m_c(1\,\rm{GeV}^2)=(1.35\pm0.10)\,\rm{GeV}$ in our works. In
Ref.\cite{Wang0903}, we also present the result with smaller  value
$m_c=1.3\,\rm{GeV}$, which can move down the central value about
$0.06\,\rm{GeV}$. The central value $M_Y=4.37\,\rm{GeV}$ is still
larger than the $D_s^\ast {\bar D}_s^\ast$ threshold about
$150\,\rm{MeV}$.

We can interpolate the scalar molecular states which consist of the
scalar, pseudoscalar, vector, axial-vector and tensor meson pairs
with the  quark currents $\bar{Q}q \bar{q}'Q$, $\bar{Q}i\gamma_5q
\bar{q}'i\gamma_5Q$, $\bar{Q}\gamma^\mu q \bar{q}'\gamma_\mu Q$,
$\bar{Q}\gamma^\mu\gamma_5q \bar{q}'\gamma_\mu\gamma_5 Q$ and
$\bar{Q}\sigma^{\mu\nu}q\bar{q}'\sigma_{\mu\nu} Q$,  respectively.
Those molecule type interpolating currents  relate with the
diquark-antidiquark  type interpolating currents through  Fierz
reordering in both the Dirac spinor space and the color space,
\begin{eqnarray}
    \left(\begin{array}{c}
             \bar{Q}q \bar{q}'Q \\[1mm]
             \bar{Q}i\gamma_5q \bar{q}'i\gamma_5Q \\[1mm]
             \bar{Q}\gamma^\mu q \bar{q}'\gamma_\mu Q \\[1mm]
             \bar{Q}\gamma^\mu\gamma_5q \bar{q}'\gamma_\mu\gamma_5 Q \\[1mm]
             \bar{Q}\sigma^{\mu\nu}q\bar{q}'\sigma_{\mu\nu} Q
    \end{array}\right)\;=\;
    \left(\begin{array}{ccccc}
    -\frac{1}{8}&  -\frac{1}{8} &\phantom{-}\frac{1}{8}&  -\frac{1}{8} &  -\frac{1}{16}    \\[1mm]
     \phantom{-}\frac{1}{8} &\phantom{-}\frac{1}{8}&\phantom{-}\frac{1}{8}&  -\frac{1}{8} &\phantom{-}\frac{1}{16}\\[1mm]
      -\frac{1}{2}      &  \phantom{-} \frac{1}{2}       &  -\frac{1}{4} &  -\frac{1}{4} &     \phantom{-} 0    \\[1mm]
       - \frac{1}{2}      &  \phantom{-} \frac{1}{2}       &\phantom{-}\frac{1}{4}&\phantom{-}\frac{1}{4}&     \phantom{-}0    \\[1mm]
          \phantom{-} \frac{3}{2}      &  \phantom{-} \frac{3}{2}       &    \phantom{-} 0     &    \phantom{-}  0    &  -\frac{1}{4}
    \end{array}\right)\;
    \left(\begin{array}{c}
             \bar{Q}\gamma_5C \lambda^a\bar{q}'Q C\gamma_5\lambda^a q \\[1mm]
             \bar{Q}C\lambda^a\bar{q}'Q C \lambda^a q \\[1mm]
            \bar{Q}\gamma^\mu\gamma_5 C\lambda^a\bar{q}'Q C\gamma_\mu\gamma_5\lambda^a q \\[1mm]
             \bar{Q}\gamma^\mu C\lambda^a\bar{q}'Q C\gamma_\mu\lambda^a q \\[1mm]
             \bar{Q}\sigma^{\mu\nu} C\lambda^a\bar{q}'Q C\sigma_{\mu\nu}\lambda^a q
    \end{array}\right) \, ,
\end{eqnarray}
where $\lambda^0=\sqrt{\frac{2}{3}}I$, the $\lambda^a$ with
$a=1,2,\cdots, 8$ are the  Gell-Mann matrixes. The $\lambda^A$ with
$A=2,5,7$ are anti-symmetric and  the $\lambda^S$ with
$S=0,1,3,4,6,8$ are symmetric.

We usually take the diquarks as the basic constituents following
Jaffe and Wilczek \cite{Jaffe2003,Jaffe2004} to construct the
tetraquark states with the diquark and antidiquark pairs.  The
diquarks have five Dirac tensor structures, scalar $C\gamma_5$,
pseudoscalar $C$, vector $C\gamma_\mu \gamma_5$, axial vector
$C\gamma_\mu $  and tensor $C\sigma_{\mu\nu}$, where  $C$ is the
charge conjunction matrix. The structures $C\gamma_\mu $ and
$C\sigma_{\mu\nu}$ are symmetric, the structures $C\gamma_5$, $C$
and $C\gamma_\mu \gamma_5$ are antisymmetric. The attractive
interactions of one-gluon exchange favor  formation of the diquarks
in  color antitriplet $\overline{3}_{ c}$, flavor antitriplet
$\overline{3}_{ f}$ and spin singlet $1_s$ \cite{GI1,GI2}.

Naively, we expect  the scalar tetraquark states with the structures
$C\gamma_5\lambda^A-\gamma_5C\lambda^A$  and $C\lambda^A-C\lambda^A$
 have the smallest masses. In
Refs.\cite{Wang09PRD,Wang09JPG},
 we study the scalar and
vector hidden charm and hidden bottom tetraquark states which
consist of $C\gamma_5\lambda^A-\gamma_5C\lambda^A$  type and
$C\gamma_\mu\lambda^A-C\lambda^A$ (and
$C\gamma_\mu\gamma_5\lambda^A-\gamma_5C\lambda^A$) type diquark
pairs respectively in a systematic way; and observe that the masses
of the vector tetraquark states are about $(0.6-0.7)\,\rm{GeV}$
larger than the corresponding ones of the scalar tetraquark states.
Furthermore, we observe that the scalar tetraquark states with the
structure $C\gamma_5\lambda^A-\gamma_5C\lambda^A$ have much smaller
masses than the corresponding ones with the structure
$C\lambda^A-C\lambda^A$ \cite{Wang09PRD}\footnote{The results with
the structure
 $C\lambda^A-C\lambda^A$  will be presented  elsewhere. }.
From Eq.(17), we  draw the conclusion tentatively that the
$\bar{Q}\gamma^\mu q \bar{q}'\gamma_\mu Q$,
$\bar{Q}\gamma^\mu\gamma_5q \bar{q}'\gamma_\mu\gamma_5 Q$,
$\bar{Q}\sigma^{\mu\nu}q\bar{q}'\sigma_{\mu\nu} Q$ type molecular
states may have smaller masses than the corresponding
 $\bar{Q}q
\bar{q}'Q$, $\bar{Q}i\gamma_5q \bar{q}'i\gamma_5Q$ type molecular
states. The conclusion is not robust enough, detailed analysis with
the QCD sum rules is still needed.

The LHCb is a dedicated $b$ and $c$-physics precision experiment at
the LHC (large hadron collider). The LHC will be the world's most
copious  source of the $b$ hadrons, and  a complete spectrum of the
$b$ hadrons will be available through gluon fusion. In proton-proton
collisions at $\sqrt{s}=14\,\rm{TeV}$¡Ì, the $b\bar{b}$ cross
section is expected to be $\sim 500\mu b$ producing $10^{12}$
$b\bar{b}$ pairs in a standard  year of running at the LHCb
operational luminosity of $2\times10^{32} \rm{cm}^{-2}
\rm{sec}^{-1}$ \cite{LHC}. The scalar ${D'}^\ast {\bar {D'}}^\ast$
${D'}_s^\ast {\bar {D'}}_s^\ast$, ${B'}^\ast {\bar {B'}}^\ast$ and
${B'}_s^\ast {\bar {B'}}_s^\ast$ molecular states predicted in the
present work may be observed at the LHCb, if they exist  indeed. We
can search for the  hidden charm molecular states  in the
$D\bar{D}$, $D^*\bar{D^*}$, $D_s\bar{D_s}$, $D_s^*\bar{D_s^*}$,
$J/\psi \rho$, $J/\psi \phi$, $J/\psi \omega$, $\eta_c\pi$,
$\eta_c\eta$, $\cdots$ invariant mass distributions and search for
the scalar hidden bottom molecular  states  in the $B\bar{B}$,
$B^*\bar{B^*}$, $B_s\bar{B_s}$, $B_s^*\bar{B_s^*}$, $\Upsilon \rho$,
$\Upsilon \phi$, $\Upsilon \omega$, $\eta_b\pi$, $\eta_b\eta$,
$\cdots$ invariant mass distributions. Those decays maybe take place
through final-state re-scattering precesses with  exchanges of the
intermediate mesons $\sigma$, $\pi$, $\rho$, $D$, $D^*$, $\cdots$ in
the $t$ channels.

\begin{figure}
\centering
\includegraphics[totalheight=5cm,width=6cm]{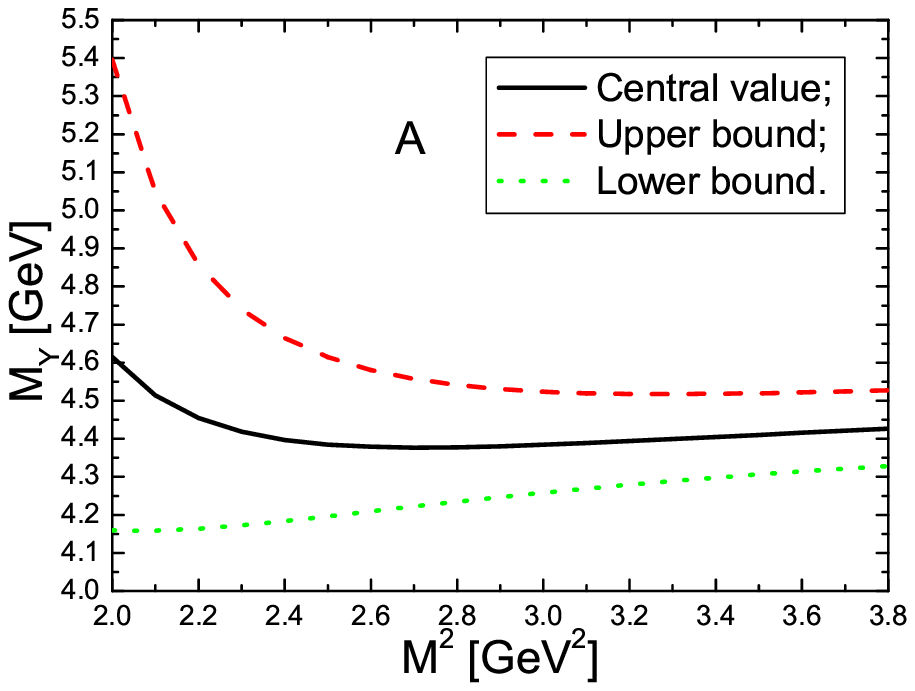}
\includegraphics[totalheight=5cm,width=6cm]{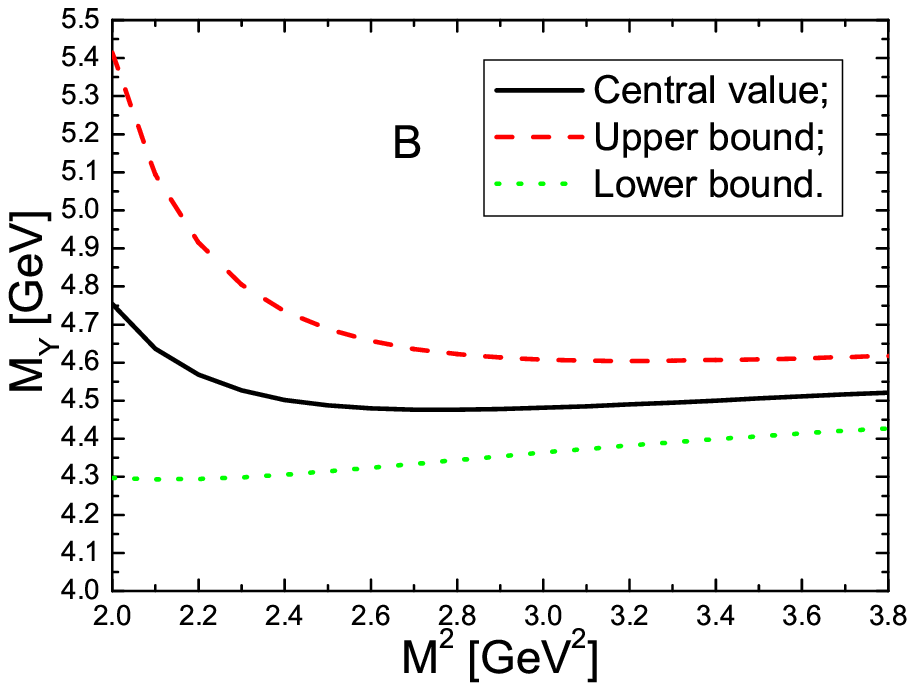}
\includegraphics[totalheight=5cm,width=6cm]{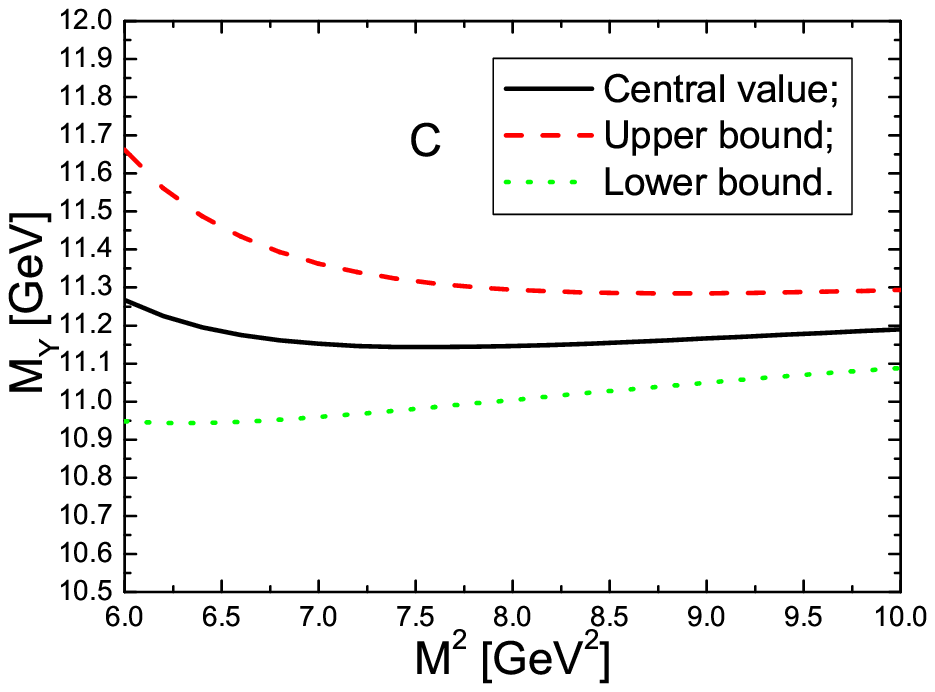}
\includegraphics[totalheight=5cm,width=6cm]{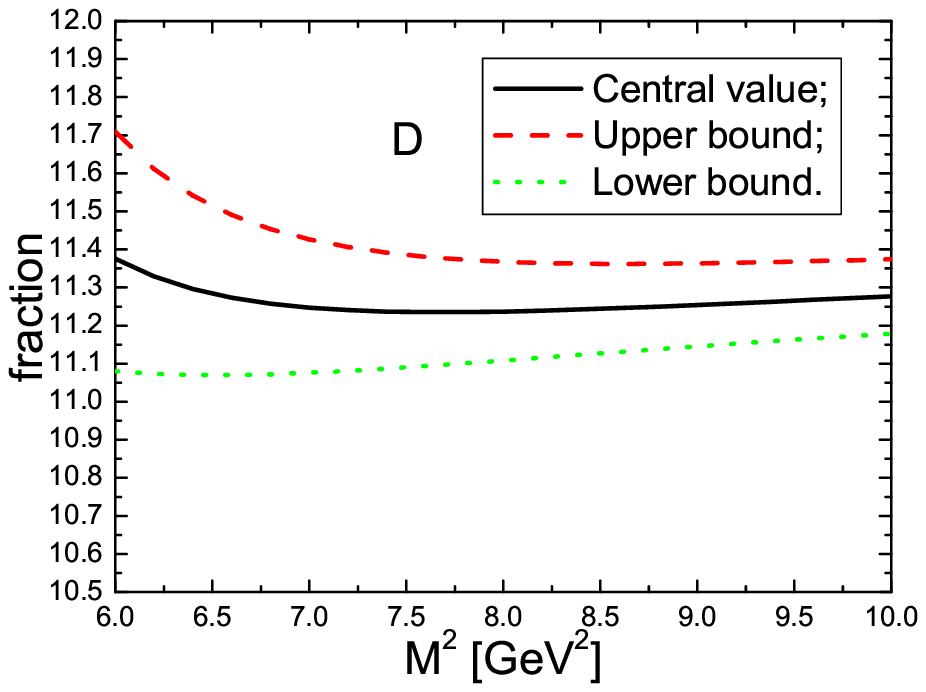}
  \caption{ The masses of the scalar molecular states  with variation of the Borel parameter $M^2$. The $A$, $B$, $C$
   and $D$ denote the $\bar{c}\gamma_\mu u \bar{d} \gamma^\mu c$,
   $\bar{c}\gamma_\mu s \bar{s} \gamma^\mu c$, $\bar{b}\gamma_\mu u \bar{d} \gamma^\mu b$,
   and $\bar{b}\gamma_\mu s \bar{s} \gamma^\mu b$  channels, respectively.   }
\end{figure}

\begin{figure}
\centering
\includegraphics[totalheight=5cm,width=6cm]{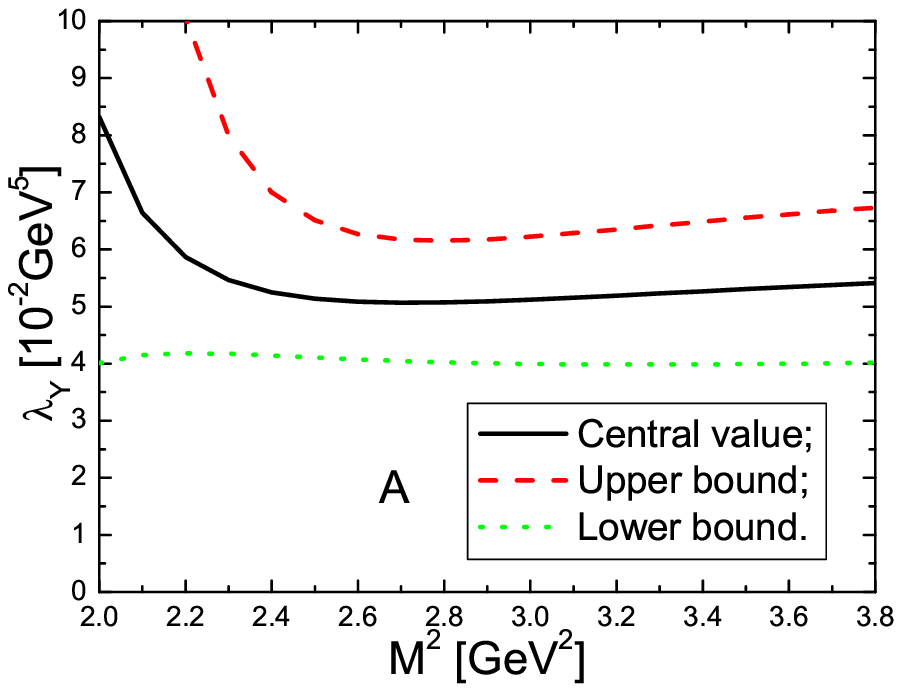}
\includegraphics[totalheight=5cm,width=6cm]{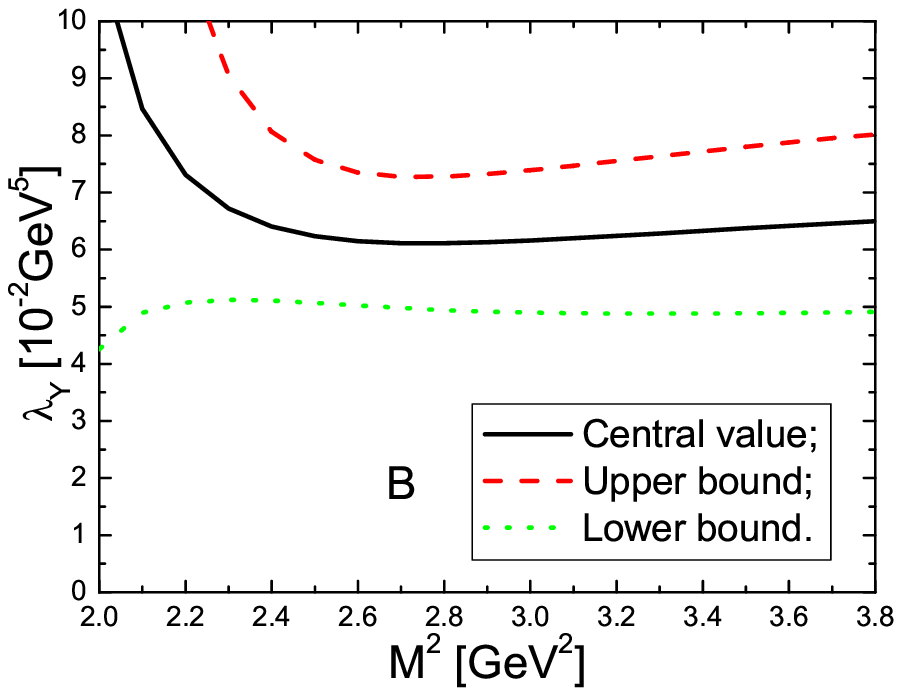}
\includegraphics[totalheight=5cm,width=6cm]{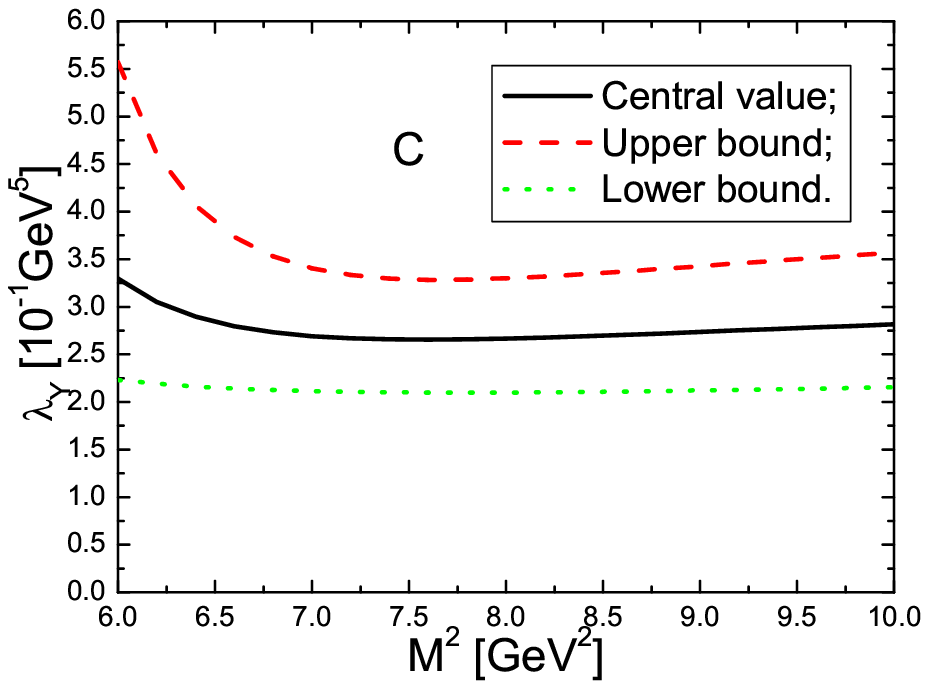}
\includegraphics[totalheight=5cm,width=6cm]{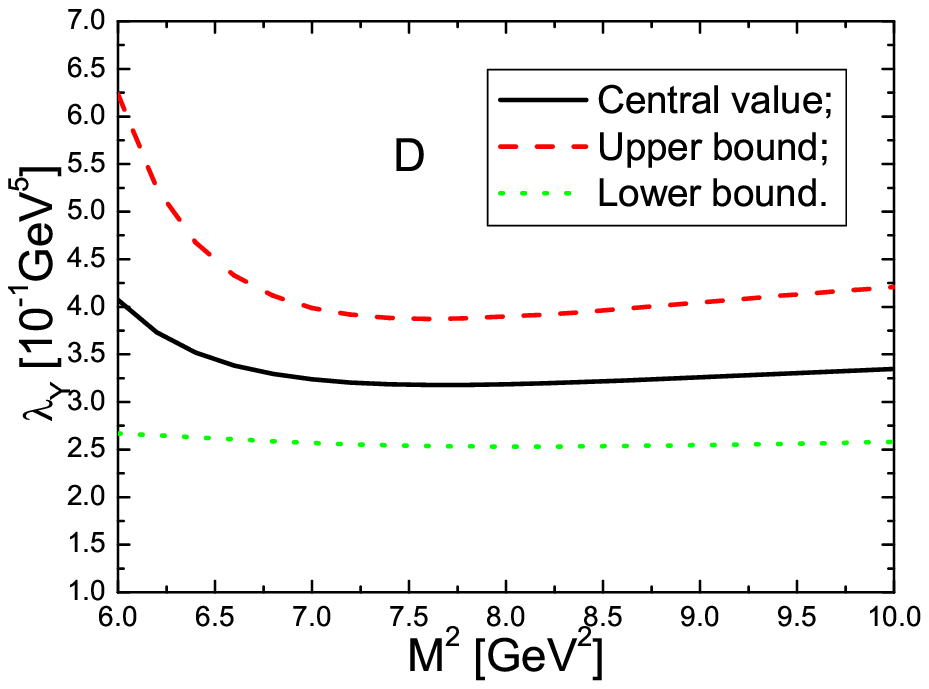}
  \caption{ The pole residues of the scalar molecular states  with variation of the Borel parameter $M^2$. The $A$, $B$,
  $C$
   and $D$ denote the $\bar{c}\gamma_\mu u \bar{d} \gamma^\mu c$,
   $\bar{c}\gamma_\mu s \bar{s} \gamma^\mu c$, $\bar{b}\gamma_\mu u \bar{d} \gamma^\mu b$,
   and $\bar{b}\gamma_\mu s \bar{s} \gamma^\mu b$  channels, respectively. }
\end{figure}

\begin{table}
\begin{center}
\begin{tabular}{|c|c|c|c|}
\hline\hline molecular states & masses & thresholds \cite{PDG}\\
\hline
      $\bar{c}\gamma_\alpha u \bar{d} \gamma^\alpha c$  &$4.38\pm0.18$& $4.014$\\ \hline
            $\bar{c}\gamma_\alpha s \bar{s} \gamma^\alpha c$ &$4.48\pm0.17$& $4.224$\\      \hline
    $\bar{b}\gamma_\alpha u \bar{d} \gamma^\alpha b$  &$11.14\pm0.19$&$10.650$ \\ \hline
            $ \bar{b}\gamma_\alpha s\bar{s} \gamma^\alpha b $ &$11.24\pm0.18$& $10.831$\\      \hline
    \hline
\end{tabular}
\end{center}
\caption{ The masses (in unit of GeV)  of the scalar molecular
states. }
\end{table}

\begin{table}
\begin{center}
\begin{tabular}{|c|c|c|}
\hline\hline molecular states & pole residues  \\
\hline
      $\bar{c}\gamma_\alpha u \bar{d} \gamma^\alpha c$  &$5.1\pm1.1$ \\ \hline
            $\bar{c}\gamma_\alpha s\bar{s} \gamma^\alpha c $ &$6.2\pm1.1$\\      \hline
    $\bar{b}\gamma_\alpha u \bar{d} \gamma^\alpha b$  &$2.7\pm0.6$ \\ \hline
            $ \bar{b}\gamma_\alpha s\bar{s} \gamma^\alpha b $ &$3.2\pm0.7$ \\      \hline
    \hline
\end{tabular}
\end{center}
\caption{ The  pole residues (in unit of $10^{-2}\, \rm{GeV}^5$ and
$10^{-1}\, \rm{GeV}^5$ for the hidden charm and  bottom channels
respectively) of the scalar molecular states. }
\end{table}

The QCD sum rules  is just a QCD-inspired model, we calculate the
ground state mass   by imposing the two criteria (pole dominance and
convergence of the operator product expansion) of the QCD sum rules.
In fact, we can take smaller threshold parameter $s_0$ and larger
Borel parameter $M^2$ to reproduce the experimental value of the
$Y(4140)$ as a scalar ${D}_s^\ast {\bar {D}}_s^\ast$ molecular state
by releasing the pole dominance condition.
 We usually consult the experimental data in choosing the
Borel parameter $M^2$ and the threshold parameter $s_0$. The present
experimental knowledge about the phenomenological hadronic spectral
densities of the multiquark states (irrespective of the molecule
type and the diquark-antidiquark type) is  rather vague. More
experimental data are still needed.

\section{Conclusion}
In this article, we assume that there exist the scalar ${D}^\ast
{\bar {D}}^\ast$, ${D}_s^\ast {\bar {D}}_s^\ast$, ${B}^\ast {\bar
{B}}^\ast$ and ${B}_s^\ast {\bar {B}}_s^\ast$ molecular states, and
study their masses using the QCD sum rules. Our predictions depend
heavily on  the two criteria (pole dominance and convergence of the
operator product expansion) of the QCD sum rules. The numerical
results indicate that the masses are about $(250-500)\,\rm{MeV}$
above the corresponding ${D}^\ast -{\bar {D}}^\ast$, ${D}_s^\ast
-{\bar {D}}_s^\ast$, ${B}^\ast -{\bar {B}}^\ast$ and ${B}_s^\ast
-{\bar {B}}_s^\ast$ thresholds, the $Y(4140)$ is unlikely  a scalar
${D}_s^\ast {\bar {D}}_s^\ast$ molecular state. The scalar $D^\ast
{\bar D}^\ast$, $D_s^\ast {\bar D}_s^\ast$, $B^\ast {\bar B}^\ast$,
$B_s^\ast {\bar B}_s^\ast$ molecular states maybe not exist, while
the scalar ${D'}^\ast {\bar {D'}}^\ast$ ${D'}_s^\ast {\bar
{D'}}_s^\ast$, ${B'}^\ast {\bar {B'}}^\ast$ and ${B'}_s^\ast {\bar
{B'}}_s^\ast$ molecular states maybe exist, and may be observed at
the LHCb.

\section*{Appendix}
The spectral densities at the level of the quark-gluon degrees of
freedom:
\begin{eqnarray}
\rho_0(s)&=&\frac{3}{1024 \pi^6}
\int_{\alpha_{i}}^{\alpha_{f}}d\alpha \int_{\beta_{i}}^{1-\alpha}
d\beta
\alpha\beta(1-\alpha-\beta)^3(s-\widetilde{m}^2_Q)^2(7s^2-6s\widetilde{m}^2_Q+\widetilde{m}^4_Q)
\nonumber \\
&&+\frac{3}{1024 \pi^6} \int_{\alpha_{i}}^{\alpha_{f}}d\alpha
\int_{\beta_{i}}^{1-\alpha} d\beta
\alpha\beta(1-\alpha-\beta)^2(s-\widetilde{m}^2_Q)^3(3s-\widetilde{m}^2_Q)
\nonumber \\
&&+\frac{3m_sm_Q}{512 \pi^6} \int_{\alpha_{i}}^{\alpha_{f}}d\alpha
\int_{\beta_{i}}^{1-\alpha} d\beta
(\alpha+\beta)(1-\alpha-\beta)^2(s-\widetilde{m}^2_Q)^2(5s-2\widetilde{m}^2_Q)
\, ,
\end{eqnarray}

\begin{eqnarray}
\rho_{\langle\bar{s}s\rangle}(s)&=&\frac{3m_s\langle\bar{s}s\rangle}{32
\pi^4} \int_{\alpha_{i}}^{\alpha_{f}}d\alpha
\int_{\beta_{i}}^{1-\alpha} d\beta
\alpha\beta(1-\alpha-\beta)(10s^2-12s\widetilde{m}^2_Q+3\widetilde{m}^4_Q)
\nonumber \\
&&+\frac{3m_s\langle\bar{s}s\rangle}{32 \pi^4}
\int_{\alpha_{i}}^{\alpha_{f}}d\alpha \int_{\beta_{i}}^{1-\alpha}
d\beta \alpha\beta (s-\widetilde{m}^2_Q)(2s-\widetilde{m}^2_Q)
\nonumber \\
&&-\frac{m_s\langle\bar{s} g_s \sigma Gs\rangle}{64 \pi^4}
\int_{\alpha_{i}}^{\alpha_{f}}d\alpha \int_{\beta_{i}}^{1-\alpha}
d\beta \alpha\beta
\left[6(2s-\widetilde{m}^2_Q)+s^2\delta(s-\widetilde{m}^2_Q)\right]
\nonumber \\
&&-\frac{3m_Q\langle\bar{s}s\rangle}{32 \pi^4}
\int_{\alpha_{i}}^{\alpha_{f}}d\alpha \int_{\beta_{i}}^{1-\alpha}
d\beta (\alpha+\beta)(1-\alpha-\beta)
(s-\widetilde{m}^2_Q)(2s-\widetilde{m}^2_Q)
\nonumber \\
&&+\frac{3m_Q\langle\bar{s}g_s \sigma Gs\rangle}{128 \pi^4}
\int_{\alpha_{i}}^{\alpha_{f}}d\alpha \int_{\beta_{i}}^{1-\alpha}
d\beta (\alpha+\beta) (3s-2\widetilde{m}^2_Q)
\nonumber \\
&&-\frac{3m_sm_Q^2\langle\bar{s}s\rangle}{8 \pi^4}
\int_{\alpha_{i}}^{\alpha_{f}}d\alpha \int_{\beta_{i}}^{1-\alpha}
d\beta
(s-\widetilde{m}^2_Q)\nonumber \\
&&-\frac{m_s\langle\bar{s}g_s \sigma Gs\rangle}{64 \pi^4}
\int_{\alpha_{i}}^{\alpha_{f}}d\alpha
 \alpha(1-\alpha) (3s-2\widetilde{\widetilde{m}}_Q^2)\nonumber \\
&&+\frac{3m_sm_Q^2\langle\bar{s}g_s \sigma Gs\rangle}{32 \pi^4}
\int_{\alpha_{i}}^{\alpha_{f}}d\alpha \, ,
\end{eqnarray}

\begin{eqnarray}
\rho_{\langle\bar{s}s\rangle^2}(s)&=&\frac{m_Q^2\langle\bar{s}s\rangle^2}{4
\pi^2} \int_{\alpha_{i}}^{\alpha_{f}}d\alpha
+\frac{m_Q^2\langle\bar{s}g_s \sigma Gs\rangle^2}{64
\pi^2M^6} \int_{\alpha_{i}}^{\alpha_{f}}d\alpha s^2 \delta(s-\widetilde{\widetilde{m}}_Q^2)\nonumber \\
&&-\frac{m_Q^2\langle\bar{s}s\rangle\langle\bar{s}g_s \sigma
Gs\rangle}{8 \pi^2} \int_{\alpha_{i}}^{\alpha_{f}}d\alpha
\left[1+\frac{s}{M^2} \right]
\delta(s-\widetilde{\widetilde{m}}_Q^2)\nonumber \\
&&-\frac{m_sm_Q\langle\bar{s}s\rangle^2}{16 \pi^2}
\int_{\alpha_{i}}^{\alpha_{f}}d\alpha
\left[2+s \delta(s-\widetilde{\widetilde{m}}_Q^2)\right]\nonumber \\
&&+\frac{5m_sm_Q\langle\bar{s}s\rangle\langle\bar{s}g_s \sigma
Gs\rangle}{96\pi^2} \int_{\alpha_{i}}^{\alpha_{f}}d\alpha
\left[1+\frac{s}{M^2}+\frac{s^2}{2M^4}
\right]\delta(s-\widetilde{\widetilde{m}}_Q^2) \, ,
\end{eqnarray}

\begin{eqnarray}
\rho^A_{\langle GG\rangle}(s)&=&-\frac{m_Q^2}{256 \pi^4}
\int_{\alpha_{i}}^{\alpha_{f}}d\alpha \int_{\beta_{i}}^{1-\alpha}
d\beta \left(\frac{\alpha}{\beta^2}+\frac{\beta}{\alpha^2}
\right)(1-\alpha-\beta)^3\left[2s-\widetilde{m}^2_Q+\frac{s^2}{6}\delta(s-\widetilde{m}^2_Q)\right]
\nonumber \\
&&+\frac{3m_sm_Q-m_Q^2}{512 \pi^4}
\int_{\alpha_{i}}^{\alpha_{f}}d\alpha \int_{\beta_{i}}^{1-\alpha}
d\beta \left(\frac{\alpha}{\beta^2}+\frac{\beta}{\alpha^2}
\right)(1-\alpha-\beta)^2(3s-2\widetilde{m}^2_Q)\nonumber \\
&&-\frac{m_sm_Q^3}{512 \pi^4} \int_{\alpha_{i}}^{\alpha_{f}}d\alpha
\int_{\beta_{i}}^{1-\alpha} d\beta
\left(\frac{1}{\alpha^3}+\frac{1}{\beta^3}
\right)(\alpha+\beta)(1-\alpha-\beta)^2\left[2
+s\delta(s-\widetilde{m}^2_Q)\right]
\nonumber \\
 &&-\frac{1}{512 \pi^4}
\int_{\alpha_{i}}^{\alpha_{f}}d\alpha \int_{\beta_{i}}^{1-\alpha}
d\beta
(\alpha+\beta)(1-\alpha-\beta)^2(10s^2-12s\widetilde{m}^2_Q+3\widetilde{m}^4_Q)
\nonumber \\
&&+\frac{1}{256 \pi^4} \int_{\alpha_{i}}^{\alpha_{f}}d\alpha
\int_{\beta_{i}}^{1-\alpha} d\beta
(\alpha+\beta)(1-\alpha-\beta)(s-\widetilde{m}^2_Q)(2s-\widetilde{m}^2_Q)
\nonumber \\
&&-\frac{3m_sm_Q}{128 \pi^4} \int_{\alpha_{i}}^{\alpha_{f}}d\alpha
\int_{\beta_{i}}^{1-\alpha} d\beta
(1-\alpha-\beta)(3s-2\widetilde{m}^2_Q)
\nonumber \\
&&-\frac{m_sm_Q^2\langle\bar{s}s\rangle}{96 \pi^2}
\int_{\alpha_{i}}^{\alpha_{f}}d\alpha \int_{\beta_{i}}^{1-\alpha}
d\beta \left(\frac{\alpha}{\beta^2}+\frac{\beta}{\alpha^2}
\right)(1-\alpha-\beta)\nonumber \\
&&\left[1+\frac{s}{M^2}+\frac{s^2}{2M^4}\right]\delta(s-\widetilde{m}^2_Q)\nonumber \\
&&-\frac{m_sm_Q^2\langle\bar{s}s\rangle}{192\pi^2}
\int_{\alpha_{i}}^{\alpha_{f}}d\alpha \int_{\beta_{i}}^{1-\alpha}
d\beta \left(\frac{\alpha}{\beta^2}+\frac{\beta}{\alpha^2}
\right)\left[1+\frac{s}{M^2}\right]\delta(s-\widetilde{m}^2_Q)\nonumber \\
&&+\frac{m_sm_Q^2\langle\bar{s}g_s \sigma Gs\rangle}{1152\pi^2M^6}
\int_{\alpha_{i}}^{\alpha_{f}}d\alpha \int_{\beta_{i}}^{1-\alpha}
d\beta \left(\frac{\alpha}{\beta^2}+\frac{\beta}{\alpha^2}
\right)\widetilde{m}^4_Q\delta(s-\widetilde{m}^2_Q)\nonumber \\
&&+\frac{m_sm_Q^4\langle\bar{s}s\rangle}{48\pi^2M^2}
\int_{\alpha_{i}}^{\alpha_{f}}d\alpha \int_{\beta_{i}}^{1-\alpha}
d\beta \left(\frac{1}{\alpha^3}+\frac{1}{\beta^3}
\right)\delta(s-\widetilde{m}^2_c)\nonumber \\
&&+\frac{m_Q^3\langle\bar{s}s\rangle}{192 \pi^2}
\int_{\alpha_{i}}^{\alpha_{f}}d\alpha \int_{\beta_{i}}^{1-\alpha}
d\beta \left(\frac{1}{\alpha^3}+\frac{1}{\beta^3}
\right)(\alpha+\beta)(1-\alpha-\beta)\nonumber\\
&&\left[1+\frac{s}{M^2} \right]\delta(s-\widetilde{m}^2_Q)
\nonumber \\
&&-\frac{m_Q^3\langle\bar{s}g_s \sigma Gs\rangle}{768 \pi^2M^4}
\int_{\alpha_{i}}^{\alpha_{f}}d\alpha \int_{\beta_{i}}^{1-\alpha}
d\beta \left(\frac{1}{\alpha^3}+\frac{1}{\beta^3}
\right)(\alpha+\beta)\widetilde{m}^2_Q\delta(s-\widetilde{m}^2_Q)
\nonumber \\
&&-\frac{m_Q\langle\bar{s}s\rangle}{64 \pi^2}
\int_{\alpha_{i}}^{\alpha_{f}}d\alpha \int_{\beta_{i}}^{1-\alpha}
d\beta \left(\frac{\alpha}{\beta^2}+\frac{\beta}{\alpha^2}
\right)(1-\alpha-\beta)\left[2+s\delta(s-\widetilde{m}^2_Q)\right]\nonumber \\
&&+\frac{m_Q\langle\bar{s}g_s \sigma Gs\rangle}{256 \pi^2}
\int_{\alpha_{i}}^{\alpha_{f}}d\alpha \int_{\beta_{i}}^{1-\alpha}
d\beta \left(\frac{\alpha}{\beta^2}+\frac{\beta}{\alpha^2}
\right)\left[1+\frac{s}{M^2}\right]\delta(s-\widetilde{m}^2_Q)\nonumber \\
&&-\frac{m_sm_Q^2\langle\bar{s}s\rangle}{16 \pi^2}
\int_{\alpha_{i}}^{\alpha_{f}}d\alpha \int_{\beta_{i}}^{1-\alpha}
d\beta \left(\frac{1}{\alpha^2}+\frac{1}{\beta^2}
\right)\delta(s-\widetilde{m}^2_Q)\nonumber \\
&&-\frac{m_s\langle\bar{s}s\rangle}{64 \pi^2}
\int_{\alpha_{i}}^{\alpha_{f}}d\alpha \int_{\beta_{i}}^{1-\alpha}
d\beta
(\alpha+\beta)\left[1+\frac{2s}{3}\delta(s-\widetilde{m}^2_Q)+\frac{s^2}{6M^2}
\delta(s-\widetilde{m}^2_Q)\right]\nonumber \\
&&+\frac{m_Q\langle\bar{s}s\rangle}{32 \pi^2}
\int_{\alpha_{i}}^{\alpha_{f}}d\alpha \int_{\beta_{i}}^{1-\alpha}
d\beta \left[2+s\delta(s-\widetilde{m}^2_Q)\right] \, ,
\end{eqnarray}

\begin{eqnarray}
\rho^B_{\langle
GG\rangle}(s)&=&-\frac{m_Q^4\langle\bar{s}s\rangle^2}{72M^4}
\int_{\alpha_{i}}^{\alpha_{f}}d\alpha
\left[\frac{1}{\alpha^3}+\frac{1}{(1-\alpha)^3}
\right]\delta(s-\widetilde{\widetilde{m}}_Q^2)\nonumber \\
&&-\frac{m_sm_Q^4\langle\bar{s}g_s \sigma Gs\rangle}{192\pi^2M^4}
\int_{\alpha_{i}}^{\alpha_{f}}d\alpha
\left[\frac{1}{\alpha^3}+\frac{1}{(1-\alpha)^3}
\right]\delta(s-\widetilde{\widetilde{m}}_Q^2)\nonumber \\
&&+\frac{m_sm_Q^2\langle\bar{s}g_s \sigma Gs\rangle}{1152\pi^2M^4}
\int_{\alpha_{i}}^{\alpha_{f}}d\alpha
\left[\frac{1-\alpha}{\alpha^2}+\frac{\alpha}{(1-\alpha)^2}
\right]\widetilde{\widetilde{m}}_Q^2\delta(s-\widetilde{\widetilde{m}}_Q^2)\nonumber \\
&&-\frac{m_sm_Q^3\langle\bar{s}s\rangle^2}{288M^4}
\int_{\alpha_{i}}^{\alpha_{f}}d\alpha
\left[\frac{1}{\alpha^3}+\frac{1}{(1-\alpha)^3}
\right]\left[1-\frac{s}{M^2}\right]\delta(s-\widetilde{\widetilde{m}}_Q^2)\nonumber \\
&&+\frac{m_Q^2\langle\bar{s}s\rangle^2}{24M^2}
\int_{\alpha_{i}}^{\alpha_{f}}d\alpha
\left[\frac{1}{\alpha^2}+\frac{1}{(1-\alpha)^2}
\right]\delta(s-\widetilde{\widetilde{m}}_Q^2)\nonumber \\
&&+\frac{m_sm_Q^2\langle\bar{s}g_s \sigma Gs\rangle}{64\pi^2M^2}
\int_{\alpha_{i}}^{\alpha_{f}}d\alpha
\left[\frac{1}{\alpha^2}+\frac{1}{(1-\alpha)^2}
\right]\delta(s-\widetilde{\widetilde{m}}_Q^2)\nonumber \\
&&-\frac{m_sm_Q\langle\bar{s}s\rangle^2}{96M^4}
\int_{\alpha_{i}}^{\alpha_{f}}d\alpha
\left[\frac{1-\alpha}{\alpha^2}+\frac{\alpha}{(1-\alpha)^2}
\right]\widetilde{\widetilde{m}}_Q^2\delta(s-\widetilde{\widetilde{m}}_Q^2)\nonumber \\
&&+\frac{m_s\langle\bar{s}s\rangle}{384\pi^2}
\int_{\alpha_{i}}^{\alpha_{f}}d\alpha
\left[2+s\delta(s-\widetilde{\widetilde{m}}_Q^2)
\right]\nonumber \\
&&-\frac{m_Q\langle\bar{s}g_s \sigma Gs\rangle}{128\pi^2}
\int_{\alpha_{i}}^{\alpha_{f}}d\alpha
\left[1+\frac{s}{M^2}\right]\delta(s-\widetilde{\widetilde{m}}_Q^2)
\, ,
\end{eqnarray}
where $\alpha_{f}=\frac{1+\sqrt{1-\frac{4m_Q^2}{s}}}{2}$,
$\alpha_{i}=\frac{1-\sqrt{1-\frac{4m_Q^2}{s}}}{2}$,
$\beta_{i}=\frac{\alpha m_Q^2}{\alpha s -m_Q^2}$,
$\widetilde{m}_Q^2=\frac{(\alpha+\beta)m_Q^2}{\alpha\beta}$,
$\widetilde{\widetilde{m}}_Q^2=\frac{m_Q^2}{\alpha(1-\alpha)}$.

\section*{Acknowledgements}
This  work is supported by National Natural Science Foundation of
China, Grant Number 10775051, and Program for New Century Excellent
Talents in University, Grant Number NCET-07-0282.

\end{document}